\documentclass[iop,apj]{emulateapj}
\usepackage{apjfonts}

\gdef\kms{km\,s$^{-1}$}
\gdef\msun{$M_{\odot}$}
\gdef\targ{NGC\,4472}
\lefthead{van Dokkum et al.}
\righthead{}
\slugcomment{Accepted for publication in the Astrophysical Journal}
\begin{document}

\title{Fluctuation Spectroscopy: A New Probe 
of Old Stellar Populations}

\author{Pieter G.\ van Dokkum\altaffilmark{1} and
Charlie Conroy\altaffilmark{2,3}
}

\altaffiltext{1}
{Department of Astronomy, Yale University, New Haven, CT 06511, USA}
\altaffiltext{2}
{Department of Astronomy \& Astrophysics, University of California, Santa Cruz, CA, USA}
\altaffiltext{3}
{Harvard-Smithsonian Center for Astrophysics, 60 Garden St., Cambridge, MA, USA}

\begin{abstract}

We introduce a new method to determine the relative contributions
of different types of stars to the integrated light of nearby early-type
galaxies. As is well known,
the surface brightness of these galaxies shows
pixel-to-pixel fluctuations
due to Poisson variations in the number of
giant stars.
Differential spectroscopy of
pixels as a function of fluctuation strength (``fluctuation spectroscopy'')
effectively measures the
spectral variation of stars as a function of their
luminosity, information
that is otherwise difficult
to obtain for individual stars outside of
the Local Group. We apply this technique to the elliptical
galaxy \targ, using HST/ACS imaging 
in six narrow-band ramp filters tuned to spectral features
in the range 0.8\,$\mu$m -- 1.0\,$\mu$m. Pixels with
$\pm 5$\% broad-band variations show
differential color variations of 0.1\% -- 1.0\% in the narrow-band filters.
These variations are primarily due to the systematic increase
in TiO absorption strength with
increasing luminosity on
the upper giant branch. The data
are very well reproduced by
the same Conroy \& van Dokkum (2012) stellar population synthesis model
that is the best fit to the integrated light,
with residuals in the range 0.03\% -- 0.09\%. Models with ages or
metallicities that are significantly
different from the integrated-light values
do not yield good fits.
We can also rule out several modifications to the underlying model,
including the presence of a significant ($>3$\,\% of the light)
population of late M giants.
The current observations constitute a powerful test of the expected
luminosities and temperatures of metal-rich giants in
massive early-type
galaxies. Studies of pixels
with much larger (negative) fluctuations will provide unique information on
main sequence stars and the stellar initial mass function.

\end{abstract}

\keywords{
galaxies: elliptical and lenticular, cD ---
galaxies: stellar content ---
stars: evolution ---
stars: AGB and post-AGB
}

\section{Introduction}

Most of our knowledge of the stellar populations of galaxies
comes from fitting stellar population synthesis models to their
integrated light, as it is prohibitively difficult to characterize
individual stars outside of the local volume.
These models are commonly used to measure
the total stellar masses of galaxies, as well as
their ages, elemental abundances, star formation
rates, dust content,
stellar initial mass function, and other properties
(see the reviews by {Walcher} {et~al.} 2011; {Conroy} 2013, and references therein).
Many of the model ingredients carry significant uncertainties,
which translate into uncertainties in derived stellar
population parameters. The uncertainties in derived masses, star formation
rates, and other parameters can be $\sim 0.3$\,dex or more
(see, e.g., {Conroy}, {Gunn}, \& {White} 2009; {Mitchell} {et~al.} 2013).

\begin{figure*}[hbtp]
\epsfxsize=16.5cm
\epsffile{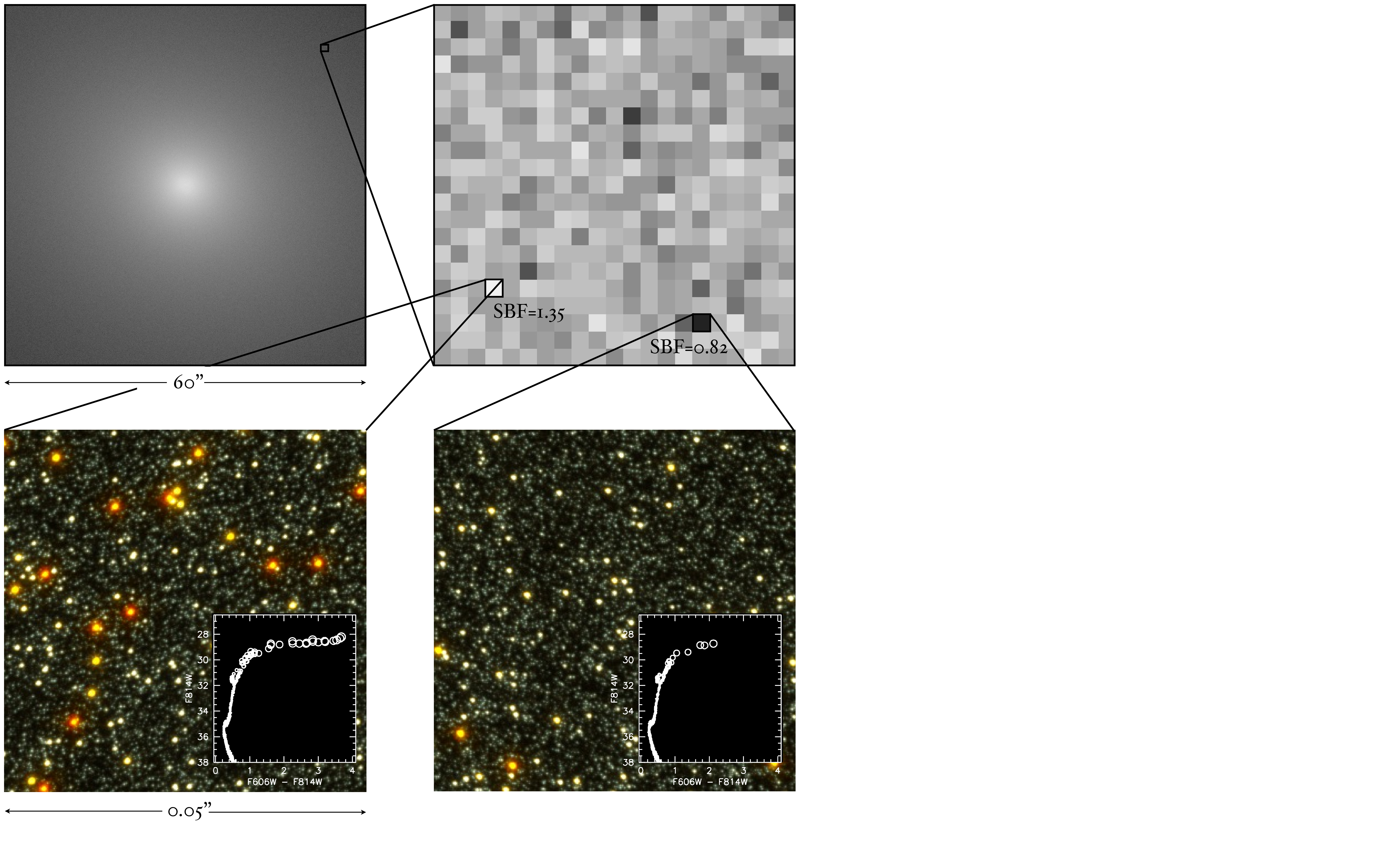}
\caption{\small
Illustration of the pixel-to-pixel variation in the stellar population
of an elliptical galaxy. The top panels show the
brightness variations in a model elliptical galaxy, based
on the light distribution of NGC\,4472, as
observed with an idealized HST + ACS
(with a PSF that is a $\delta$-function).
The bottom
panels show a bright 
and a faint pixel in the ACS image,
as observed with a hypothetical diffraction-limited
800\,m telescope. The insets
show color-magnitude diagrams of the stars in these pixels.
In faint pixels there is a deficit of luminous giants with
respect to the mean, and in bright pixels there is an excess.
\label{demo.fig}}
\end{figure*}

%A well-known example of such
%an uncertainty is the contribution of TP-AGB stars to the
%integrated light of $\sim 1$\,Gyr old stellar populations.
%{Maraston} (2005) suggested that stars in this
%luminous, short-lived phase of stellar evolution could have
%a much larger contribution to the integrated
%near-IR light than previously thought. As a consequence,
%the {Maraston} (2005) models tend to yield lower
%stellar masses of quiescent galaxies at $z=1-2$ than the
%{Bruzual} \& {Charlot} (2003) or {Conroy} {et~al.} (2009) models.

The predictions of
stellar population synthesis models have historically
been tested with star
clusters, as stellar population parameters
derived from the integrated light can be compared to those derived
from the distribution of stars in the color-magnitude diagram (CMD)
(e.g., {Worthey} 1994; {Bruzual} \& {Charlot} 2003; {Maraston} 2005; {Schiavon} 2007; {Conroy} {et~al.} 2009; {Thomas}, {Johansson}, \&  {Maraston} 2011; {No{\"e}l} {et~al.} 2013).
In fact, models are frequently tuned and calibrated to fit
available data from open clusters in the Milky Way and the Magellanic 
Clouds, as well as the ancient, and generally metal-poor,
globular clusters. Thanks to programs such as the ACS Nearby Galaxy
Survey Treasury (ANGST; Dalcanton et al.\ 2009), some tests can now
be done using individual stars in galaxies out to $\sim 4$\,Mpc
(see, e.g., Girardi et al.\ 2010).

%Some aspects
%can also be tested with integrated-light observations.
%Returning to
%the issue of TP-AGB stars, the differences between the {Bruzual} \& {Charlot} (2003)
%and {Maraston} (2005) models are sufficiently dramatic that they
%are detectable in galaxies whose light is dominated by $\sim 1$\,Gyr
%old stellar populations. {Conroy} \& {Gunn} (2010),
%{Kriek} {et~al.} (2010), and {Zibetti} {et~al.} (2013) have
%shown that the SEDs and spectra of such galaxies 
%are inconsistent with a large contribution of TP-AGB
%stars to the integrated near-IR light.

Despite advances like these,
the performance of the models remains essentially untested
in important regions of parameter space. Among the most important
regimes is that of the metal-rich,
old stellar populations which
dominate the light of massive early-type galaxies. 
Correctly 
modeling these stellar populations is critical for many aspects
of astrophysics; examples are the
high mass end of the galaxy
mass function ({Croton} {et~al.} 2006)
and the core mass function (van Dokkum et al.\ 2014),
the masses of central black holes
({van der Marel} {et~al.} 1998),
the production of
atomic elements in the Universe ({Worthey}, {Faber}, \&  {Gonzalez} 1992; {Conroy}, {Graves}, \& {van  Dokkum} 2014), 
the stellar initial mass function ({van Dokkum} \& {Conroy} 2010; {Conroy} \& {van Dokkum} 2012b),
and
the physical
conditions in the densest star forming regions in the early Universe
({Krumholz} 2011; Nelson et al.\ 2014).

Unfortunately,
star clusters in the Local Group with similar
ages and abundance patterns as giant ellipticals
are rare (e.g., {Caldwell} {et~al.} 2011; {van Dokkum} \& {Conroy} 2011).
We therefore largely
rely on a combination of
Milky Way stars and theoretical models to fit the spectra
of these galaxies. Of particular concern are the red
giants that dominate the near-IR light.
The spectra of these stars have strong and ubiquitous molecular
bands of titanium-oxide (TiO) and H$_2$O, which often coincide
with spectral features that are used to constrain abundances,
ages, and the IMF 
(see, e.g., {Lan{\c c}on} \& {Wood} 2000; {Rayner}, {Cushing}, \& {Vacca} 2009).

In this paper we introduce 
a new method to test the predictions of stellar population
synthesis models for old, unresolved stellar populations.
The method makes use of the fact that giants are sufficiently rare and
luminous that Poisson fluctuations in their number
cause pixel-to-pixel fluctuations in the
surface brightness distribution of nearby
early-type galaxies. 
The existence of these
surface brightness fluctuations is well known, and they have been
used extensively to measure distances to galaxies: if a galaxy
is closer, there are less giant stars in each pixel, and the pixel-to-pixel
Poisson fluctuations around the mean will therefore be larger
(e.g., {Tonry} \& {Schneider} 1988; {Tonry} {et~al.} 2001; {Mei} {et~al.} 2005; {Blakeslee} {et~al.} 2010).

Pixels with high and low fluctuations allow us to do an
experiment that gives unique information:
we can study how the integrated spectrum changes
when giants are added or subtracted to the galaxy.
This
is illustrated in Fig.\ \ref{demo.fig}, which shows the different
stellar content of
pixels with low and high fluctuations.
%The galaxy
%image is a model that is based on the actual light distribution
%of \targ,
%and the bottom panels are based on the models of \S\,\ref{model.sec}.
The bright pixel at bottom left has many more luminous
giants than the faint pixel at bottom right; therefore,
comparing spectra of pixels with different brightness enables
us, in principle, to isolate  spectral features of
stars  on the upper giant branch. 
Furthermore, in the
regime of very large fluctuations there will be pixels (or
more appropriately, lines of sight through the galaxy) whose light
has only a small contribution from luminous
stars on the RGB and AGB.
Spectroscopy of these
pixels can provide information on the integrated light of
stars that are still on the main sequence, which
is difficult to access in any other way.

This method builds on
previous studies that compared the fluctuation amplitude in
various broad-band filters to
predictions from stellar population synthesis models
(e.g., {Worthey} 1994; {Vazdekis} {et~al.} 1997; {Blakeslee}, {Vazdekis}, \&  {Ajhar} 2001; {Liu}, {Graham}, \& {Charlot} 2002; {Jensen} {et~al.} 2003). 
A central conceptual difference with previous work is that we do not use
the absolute fluctuation signal in the analysis, which is difficult
to measure (as it requires an accurate distance, a thorough
knowledge of the effects of the PSF, and a careful treatment
of contaminating objects). Instead, the method relies
on the relative change in the spectrum $\Delta F(\lambda)/\langle
F(\lambda)\rangle$ for a given relative change in the total
flux $\Delta F/\langle F\rangle$. This is a key strength of
the method: the distance, PSF, and other effects that determine
the absolute amplitude of the fluctuations are only relevant in
that they set the dynamic range that is probed with the
data. A second difference is that we identify the fluctuations of
individual pixels rather than their statistical ensemble.

Here we take a first step toward fluctuation spectroscopy by using
imaging through the narrow-band ramp filters in the Advanced
Camera for Surveys (ACS) of HST. We study a regime where the
observed fluctuations are dominated by systematic changes in the
TiO absorption strength with temperature on the upper giant branch.
The target is \targ, the most luminous galaxy in the Virgo cluster (with
the possible exception of M87 -- depending on the chosen aperture).
The main goal of the present paper is to introduce fluctuation
spectroscopy as a method to dissect unresolved stellar populations.
We also constrain certain aspects of the stellar synthesis
model, and we show that more complete tests are possible with future
data sets.
The paper is organized as follows. In \S\,\ref{obs.sec} the HST
observations are described, with emphasis on the
geometric constraints
imposed by the ramp filters. In \S\,\ref{analysis.sec} the
variation in the ramp filters is measured as a function of the variation
in the broad-band flux of a pixel. In \S\,\ref{model.sec} we interpret
the observed variation in the ramp filters, both in the context of
a default isochrone and default population synthesis model and
in the  context of several variations of this model.
The results are summarized in \S\,\ref{conclusions.sec}.

\section{Data}
\label{obs.sec}
\subsection{Observing Strategy}

\targ\ was observed for a total of 15 HST orbits in the period Feb 22 --
March 1, 2013, in the context of program ID 12523.
We used the  ramp filters in the ACS camera to obtain images
of the galaxy at six different wavelengths. The ramp filters
transmit in a narrow optical bandpass, whose central wavelength
can be tuned by rotating the filter. Each filter consists of
three segments, which
are projected on different parts of the
WFC detector array. The choice of segment (and therefore the location
of the image on the detector) depends on the desired wavelength: within
each segment only a limited range of wavelengths can be chosen.\footnote{The
geometry of the ramp filter unit, and a more extensive description of
their properties, may be found in the ACS Handbook, in particular
\S\,7.7.2 and Fig.\ 7.5.}

\begin{figure}[hbtp]
\epsfxsize=8.5cm
\epsffile[27 173 568 690]{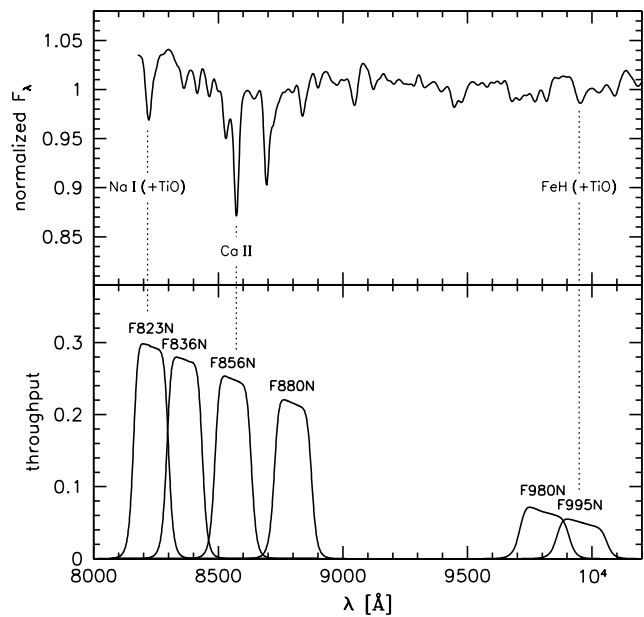}
\caption{\small
Top panel: Model spectrum of a 13 Gyr old stellar population,
redshifted to the velocity of \targ\ (997\,\kms) and smoothed
to $\sigma = 300$\,\kms. The Na\,I $\lambda 8183,8195$ doublet,
the strongest Ca triplet line Ca\,II $\lambda 8544$, and
the Wing-Ford FeH $\lambda 9916$ band are marked.
Bottom panel: The six narrow-band ramp filters that are
used in this study. We tuned three filters to the wavelengths
of important (redshifted) absorption features, with the other
three serving as
off-band filters adjacent to the primary filters.
As shown in \S\,\ref{model.sec}
the observed pixel-to-pixel variation in these six filters
is dominated by variation in the strength of TiO.
\label{filters.fig}}
\end{figure}

The six filters we used are listed in Table \ref{obs.tab}. Their
transmission curves, calculated with the {\tt synphot} package,
are shown in Fig.\ \ref{filters.fig}. The widths of the filters range
from $140$\,\AA\ for the bluest filter to $200$\,\AA\ for
the reddest, corresponding to a (sparsely sampled)
spectral resolution ranging from
$R\approx 58$ to $R\approx 50$.
The transmission curves
are compared to a model
spectrum of an old stellar population ({Conroy} \& {van Dokkum} 2012a),
redshifted to 997\,\kms\
and smoothed to a velocity dispersion of 300\,\kms.
The central wavelengths
were chosen to measure three redshifted spectral features
that are important indicators of the IMF: the Na\,I $\lambda 8183,8195$
doublet, the strongest Calcium triplet line Ca\,II $\lambda 8544$,
and the FeH $\lambda 9916$ Wing-Ford
band (e.g., {van Dokkum} \& {Conroy} 2011). For each of these features we observed
an ``on'' band and an ``off'' band, leading to six narrow-band filters.
However, as we show later, the observed pixel-to-pixel
variation in these bands in our
data is actually {\em not} caused by variation in these absorption
features, but by systematic changes in the
TiO absorption strength on the upper giant
branch (see \S\,\ref{model.sec}). To avoid confusion we do not
refer to these filters as, for example, ``Na\,I on'' and ``Na\,I off'', even
though that was how they were designed; instead, we use names
that simply refer to the central wavelength (see Table
\ref{obs.tab}).
Four of the wavelengths fall in the FR853N bandpass, which is part of the
inner segment of its ramp filter unit. These images are therefore
projected onto the IRAMP aperture, which is in the top left corner of the
WFC array. The longest wavelengths fall in the FR1016N bandpass, which
is part of the outer segment. The associated
ORAMP aperture is in the bottom right corner of the array.

\begin{table}[h]
\caption{HST ACS observations of NGC\,4472.\label{obs.tab}}
\centering
\begin{tabular}{lcclcc}
\hline\hline
Filter & $\lambda_{\rm ramp}$ [\AA] & Aperture$^a$ & Name$^b$ & Orbits & $t_{\rm exp}$ [s] \\
\hline
FR853N & 8230 & IRAMP & F823N & 2 & 5237 \\
FR853N & 8364 & IRAMP & F836N & 2 & 5237 \\
FR853N & 8558 & IRAMP & F856N & 1 & 2206 \\
FR853N & 8796 & IRAMP & F880N & 1 & 2164 \\
FR1016N & 9802 & ORAMP & F980N & 2 & 5237 \\
FR1016N & 9951 & ORAMP & F995N & 3 & 7636 \\
F814W & --- & IRAMP & F814W & 2 & 5237 \\
F814W & --- & ORAMP & F814W & 2 & 5237 \\
\hline
\end{tabular}
\end{table}
\vspace{-0.5cm}
\noindent
{\footnotesize $^{a}$\,{The aperture keyword specifies the pointing origin
that is used by HST. Depending on the wavelength, the inner (``IRAMP'')
or outer segment (``ORAMP'') of the ramp filter unit has to be used.
The galaxy is imaged on opposing corners of the WFC detector array for these
two apertures.}\\
$^b$\,{The name of the filter as it is used in the
text.}}

\begin{figure*}[hbtp]
\epsfxsize=17.8cm
\epsffile[0 125 742 487]{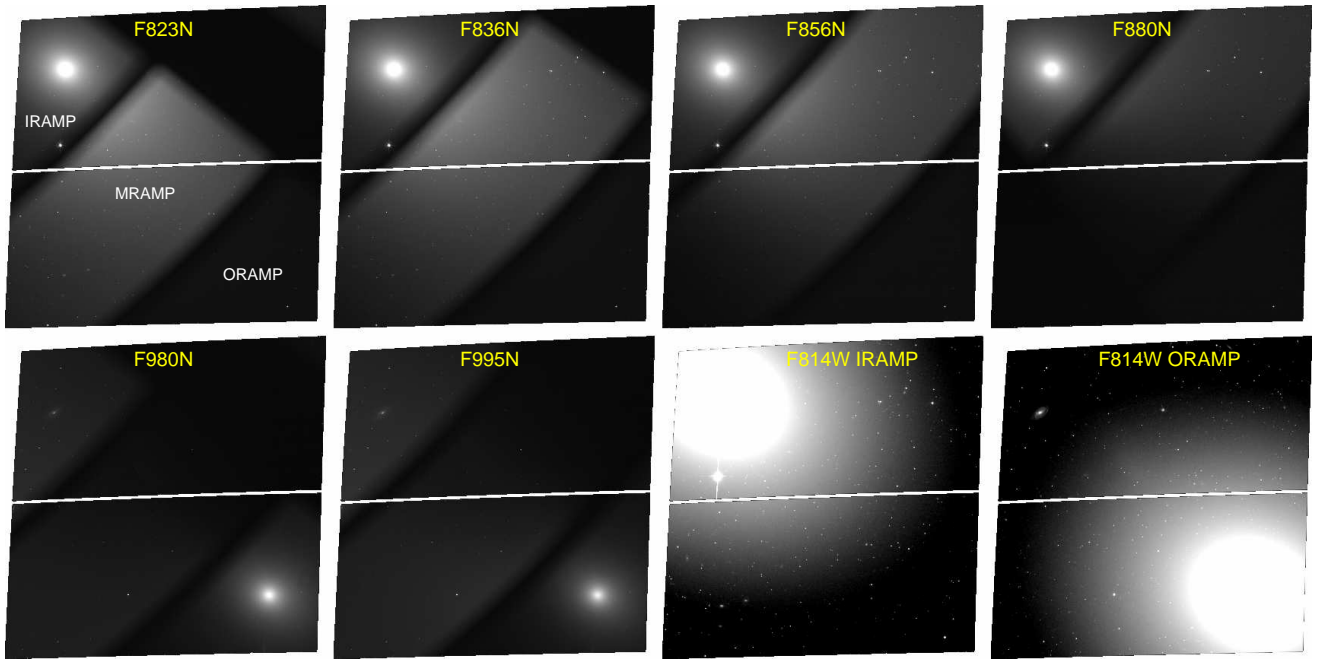}
\caption{\small
ACS images of \targ\ in the six ramp filters and in the broad
F814W filter. Four of the narrow-band filters require
the inner part of the
ramp filter structure (``IRAMP'') and two require the
outer part (``ORAMP''). A deep, 2-orbit F814W image was
obtained for each of these two apertures.
The images are scaled to the same exposure time
so that
differences in brightness indicate differences in S/N ratio.
The high S/N F814W images are used to measure the surface
brightness fluctuation in each individual pixel.
\label{filter_im.fig}}
\end{figure*}

In addition to the narrow-band data we obtained deep broad-band images
of \targ. These broad-band images are critical as they have sufficiently
high signal-to-noise ratio (S/N) to accurately measure the flux in each
pixel, and therefore the contribution of giants to that pixel relative
to its neighboring pixels. The narrow-band data do not have sufficiently
high S/N for this measurement, but that is not needed: if it is known
which pixels have high fluctuations and which pixels have low
fluctuations, we can select the giant-rich and giant-poor pixels
in the narrow-band data and average them to increase the S/N ratio.
For this to work the broad-band data and narrow-band data have to be
astrometrically-matched with an accuracy of $0.1-0.2$ pixel.
In order to limit astrometric and resampling errors we observed
\targ\ twice in the broad-band F814W filter: once using the IRAMP
pointing origin and once using the ORAMP pointing origin. This ensures that
each narrow-band observation has a matching
broad-band observation.

\subsection{Image Registration and Matching}

Pipeline-processed, drizzled data were obtained from the MAST
server.\footnote{http://archive.stsci.edu/hst/} The charge
transfer efficiency (CTE) corrected {\tt \_drc} files were used. CTE
effects are relatively small in our data as
the background is relatively high; nevertheless, the CTE correction
removes some faint streaks in the background
and produces visibly cleaner images. The data were taken in nine
visits: two visits for F995N and one visit for all other bands.

We registered the F823N, F836N, F856N, F880N, and the IRAMP F814W images
to a common reference frame. The shifts between these images are small
($\lesssim 10$\,pixels), as they all use the same pointing origin.
Globular clusters and stars in the field
were used to find the shifts; no rotation or scale corrections
were needed. We used the F814W image as a reference{} frame but first
shifted it by $\Delta x=0.4$, $\Delta y = 0.6$ pixels. This non-integer
shift ensures that the broad-band and narrow-band data are interpolated
in the same way. We used a third-order polynomial for the interpolation
to the reference{} grid. The F980N, F995N, and ORAMP F814W images were
also aligned with each other, using the same procedure. For F995N the
two visits were aligned and then added.

The point spread function (PSF) in the images needs to be exactly matched,
as any differences will influence the measured signal in the ratio of two
bands. We measured the width of the PSF by fitting Gaussians to point
sources in the field in each band. It turns out that the width of
the F814W PSF
is identical (to within $<0.1$ pixel) to the F880N PSF, but that
the FWHM of the PSFs in all other bands is narrower by $\approx 0.3$
pixels. It is not clear why the F880N PSF behaves differently from the
other ramp filters. We convolved the images to the F814W PSF, and
verified that the PSFs were identical to $<0.1$ pixel after the
convolution.  Images of the eight observations (the six ramp filters and
the F814W images in each of the two apertures) are shown in Fig.\
\ref{filter_im.fig}.

\begin{figure*}[hbtp]
\epsfxsize=16.5cm
\begin{center}
\epsffile[19 65 742 546]{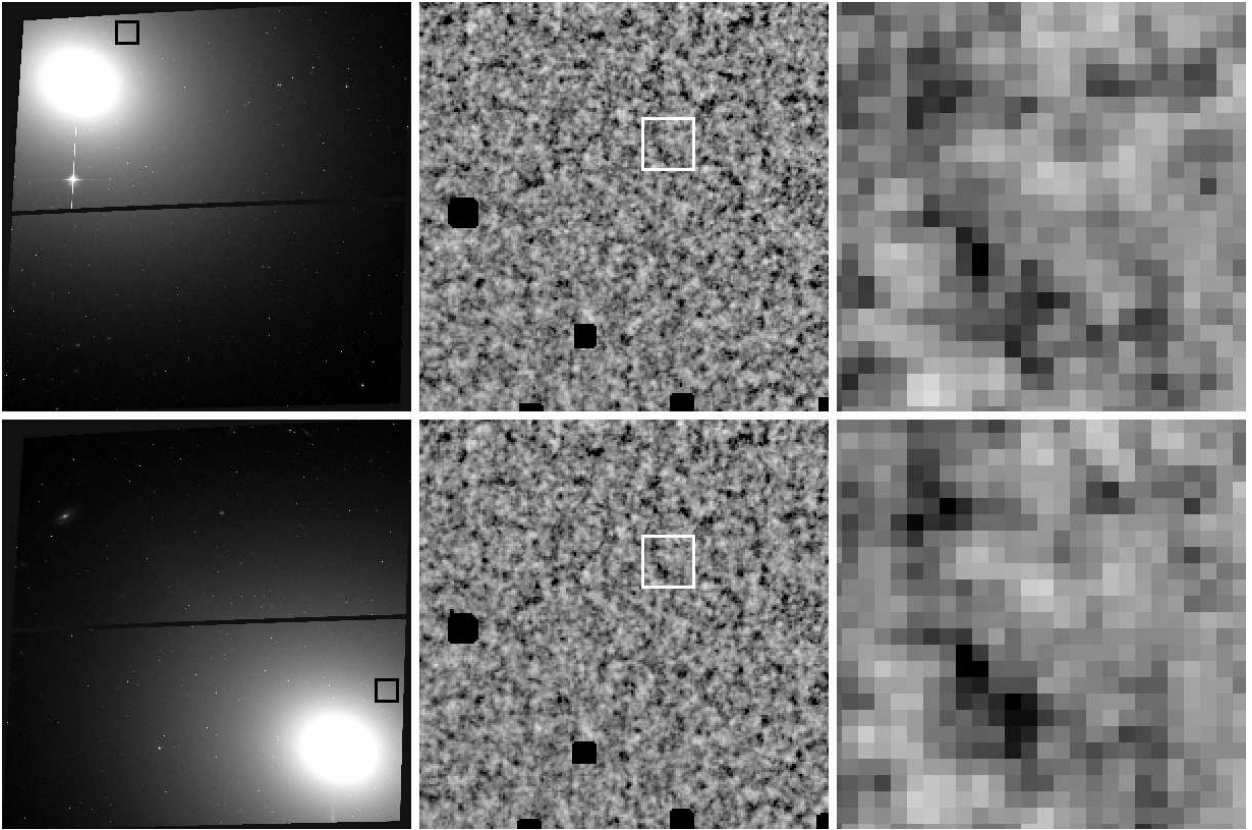}
\end{center}
\caption{\small
Surface brightness fluctuations in \targ. The left panels show the full
F814W images with \targ\ centered on the IRAMP (top) and
ORAMP (bottom) aperture locations. The middle and right panels show
surface brightness fluctuation images at different magnifications.
These images were created by dividing the
F814W images by smooth models of the galaxy light (see text).
The fluctations are not due to noise but due to variations in the number
of giants in different sightlines through the galaxy. This is demonstrated
by the fact that the fluctuations are identical in the IRAMP and ORAMP
images, which are completely independent observations.
\label{sbfdemo.fig}}
\end{figure*}

\section{Analysis}
\label{analysis.sec}

\subsection{Modeling of the Smooth Galaxy Light}

In order to determine the variation in a pixel we first need to know
what the flux in that pixel would be in the absence of fluctuations.
We used the deep F814W images to create models for the 2D light
distribution of \targ, as imaged at the IRAMP and ORAMP aperture locations.
The models
were made using the {\tt ellipse} task in the IRAF STScI package.
All parameters were left free in the fit; that is, the ellipticity,
center, and boxiness/diskiness of the ellipses were all allowed to
vary with radius.

The fits were done in two steps.
In the first step, only known bad and missing
pixels\footnote{As determined from the data quality extensions
of the fits files.} were masked in the fit. This initial fit
was subtracted from the galaxy, and strongly deviating pixels
in the residual image
($<-1.5$ or
$>1.5$ ADU\,s$^{-1}$)
were added to the bad pixel mask. These deviating
pixels are mostly at the locations of foreground stars
and globular clusters, but also include
some bad pixels that were not included in the data quality file.
In the second step the fit was improved by adding these flagged
pixels to the bad pixel mask.

\subsection{Broad-band Fluctuations of Individual Pixels}
\label{sbfmeas.sec}

The surface brightness fluctuation in each pixel was determined in the
following way. First, a conservative mask was created that includes
all pixels that might be dominated or influenced by globular clusters or
stars. In the inner regions of the galaxy these objects were
identified by dividing the image by the smooth model, as the
relative fluctuations in the center are
typically $\lesssim 1$\,\% (see below).
Pixels in the central $\sim 10\arcsec \times 10\arcsec$ that deviate by
more than 5\,\% from the smooth model were identified as potential
objects. At larger radii this is not a viable method
as the surface brightness fluctuations
approach these levels, and we identified contaminating objects
by subtracting the model from the galaxy. Away from the
central regions, pixels deviating from the
model by more than 0.5 counts/s were flagged as potential globular clusters.
Careful inspection of the residual images
showed that the combined mask correctly identifies
all ``visible'' globular clusters, stars, and background galaxies.
The method identifies the central parts of such
objects but not their faint
wings, and the final mask was created by convolving the combined inner + outer
mask with a $9\times 9$ pixel tophat.

After creating the IRAMP and ORAMP masks the F814W images
were divided by the smooth 2D models of the light distribution.
Pixels flagged in the object masks were set to
zero. The resulting residual images contain the surface brightness fluctuations
but also low level, large scale variation caused by subtle differences between
the surface brightness distribution of the galaxy and the ellipse fit models.
This large scale variation was removed using unsharp masking:
the residual images were median filtered with a $31\times 31$ pixel
mask, and divided by
these heavily smoothed versions of themselves. The quantity ``surface brightness
fluctuation'', or ``SBF'', in this paper refers to the fluxes in
these residual images. The SBF of a pixel $i$
is therefore defined as SBF\,=\,$F_{\rm F814W}^i / M_{\rm F814W}^i$,
with $M$ the flux of
a 2D model of NGC\,4472 that is constrained to fit the observed
F814W surface brightness distribution of the galaxy exactly on
$1\farcs 5$ scales.

The final surface brightness fluctuation images in the IRAMP and
ORAMP apertures are shown in Fig.\ \ref{sbfdemo.fig}. The images
show clear variation, with some pixels being brighter than the mean
and others fainter. 
The reality of
the surface brightness fluctuations is demonstrated by
their similarity in the top and bottom panels of
Fig.\ \ref{sbfdemo.fig}. The IRAMP and ORAMP observations are
entirely independent of each other: the two sets of exposures
are independent and
the galaxy is exposed on a different detector. Nevertheless, the
fluctuation pattern in the IRAMP and ORAMP data is identical, which means
that the pattern is dominated by signal, rather than noise.

\begin{figure}[hbtp]
\epsfxsize=8.6cm
\epsffile[150 306 462 456]{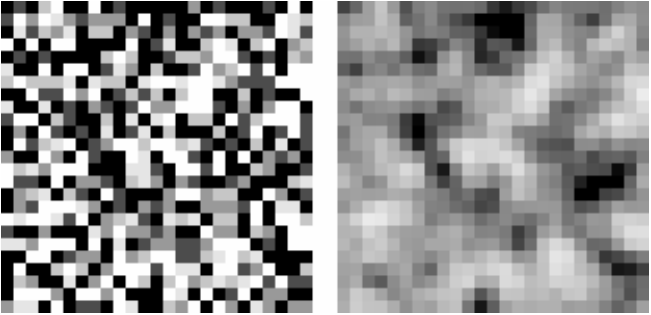}
\caption{\small
Simulation of the effects of the ACS F814W PSF on the measured
fluctuations. The left panel shows a simulated image of
uncorrelated Poisson fluctuations. The right panel shows the
same image convolved with the F814W PSF. This panel looks
similar to the right panels of Fig.\ \ref{sbfdemo.fig}; the fluctuations
are correlated, and their amplitude is reduced by a factor of 4.2
compared to the uncorrelated case.
\label{simpsf.fig}}
\end{figure}

The variation is highly correlated: clumps of
pixels vary together. This in itself shows that the pattern is
unlikely to be noise, as that is (mostly) uncorrelated in ACS data.
The correlations are caused by the HST PSF,
which smooths the intrinsically-uncorrelated variations.
This is illustrated
in Fig.\ \ref{simpsf.fig}, which shows an
image with uncorrelated Poisson noise along with the same image convolved
with the ACS F814W PSF. The correlated noise pattern is very similar
to that seen in Fig.\ \ref{sbfdemo.fig}. We find that
the amplitude of the fluctuations is reduced by a factor of 4.2;
that is, in actual ACS images the surface brightness fluctuations
are smaller by a factor of $\sim 4$ compared
to an imaginary telescope which projects delta functions on a perfect
detector with $0\farcs 05 \times 0\farcs 05$ pixels.
For reference, a Gaussian with a FWHM of 2 pixels reduces
the amplitude of random fluctuations by a factor of 3.

\begin{figure}[hbtp]
\epsfxsize=8.6cm
\epsffile[70 187 535 690]{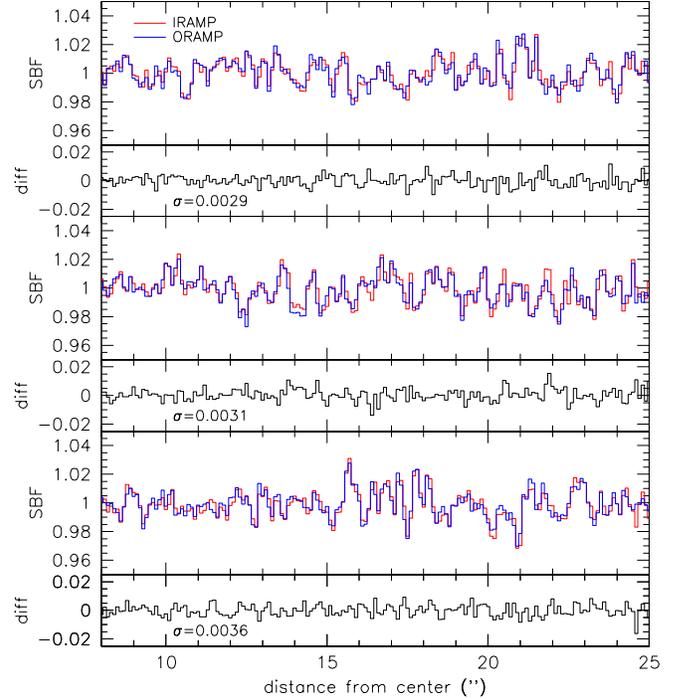}
\caption{\small
Comparison of F814W surface brightness fluctuations in the independent
IRAMP and ORAMP images. Three randomly selected 1D traces are shown,
with $2\times 4$ pixel binning. The fluctuations are very similar in the two
independent datasets, with an rms difference of 0.3\%. 
\label{traces.fig}}
\end{figure}

Although the surface brightness fluctuations in
the F814W images are dominated by signal, they are not noise-free.
We quantify the per-pixel uncertainty empirically by shifting the
IRAMP and ORAMP fluctuation images
to a common reference{} frame and subtracting them. The comparison is
done after binning the data by a factor of 2
in $x$ and a factor of 4 in $y$, to limit the effects of
sub-pixel errors in the shifts. Figure \ref{traces.fig} shows
three randomly selected 1D traces extracted from
the IRAMP and ORAMP F814W fluctuation
images. The rms scatter in the difference
is 0.003, with no strong dependence on radius in the radial
range of interest. Correcting for the binning, and taking into account
that the per-image uncertainty is smaller than the
uncertainty in the difference of two images by
a factor of $\sqrt{2}$, the per-pixel,
per-image uncertainty is $0.003\times\sqrt{2\times4}/\sqrt{2} = 0.006$.

\begin{figure*}[hbtp]
\epsfxsize=17cm
\begin{center}
\epsffile[25 503 433 687]{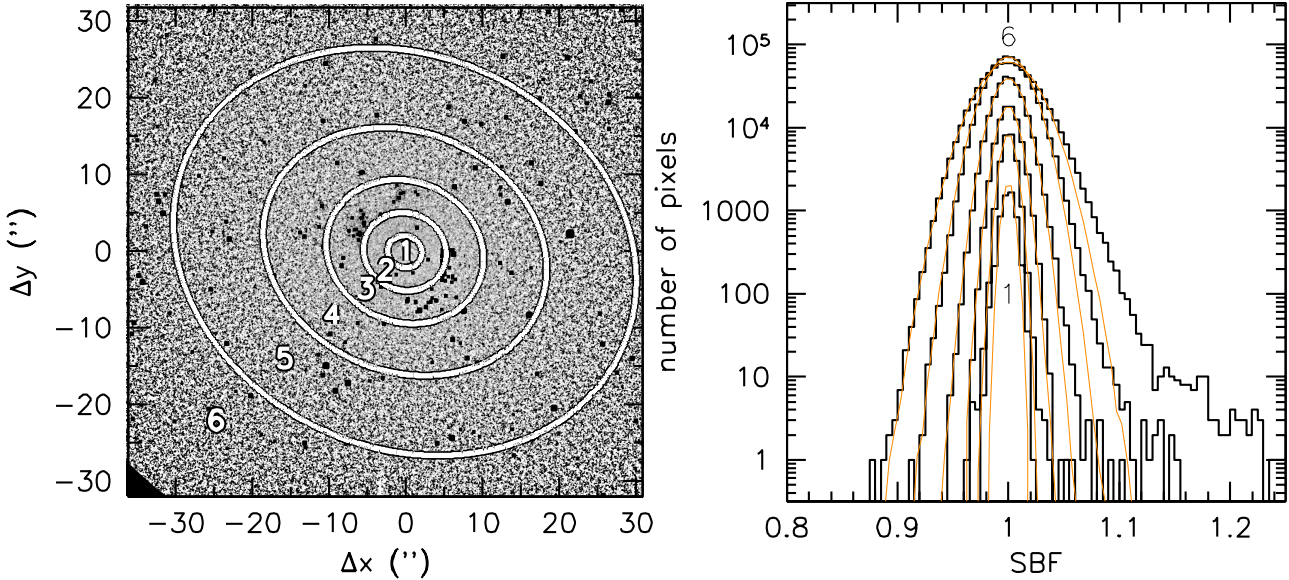}
\end{center}
\caption{\small
Left panel: broad-band surface brightness fluctuations in the F814W
ORAMP image. Ellipses indicate the galaxy model.  Right panel: measured
fluctuations in six annuli.
As expected, large fluctuations are
only found in the annulus that is furthest from the center.
Orange lines indicate the expected
amplitude of the fluctuations if they scale with $1/\sqrt{F}$.
%Going outward from the center
%the distribution is increasingly asymmetric; this reflects
%the presence of rare, very luminous giants in pixels with the largest
%fluctuations.
\label{sbf_quant.fig}}
\end{figure*}

We perform the analysis on
relatively small sections of the IRAMP and ORAMP exposures, as the
monochromatic patch is $\lesssim 1$\,arcmin  when the
ramp filter assembly is in place. The regions with uniform coverage
are approximately
$61\arcsec \times 47\arcsec$ and $67\arcsec \times 63\arcsec$
for the IRAMP and ORAMP respectively.
The ORAMP monochromatic patch
of the F814W surface brightness fluctuation
image is shown in the left panel of Fig.\ \ref{sbf_quant.fig}.
The ``mottled'' appearance is due to the surface brightness fluctuations.
The amplitude of the fluctuations increases with distance from the
galaxy center. The fluctuations are proportional to $\sqrt{N}$, with
$N$ the average number of giant stars in a pixel, which means
that the absolute
amplitude of the fluctuations is proportional to the square root
of the surface brightness. However,
the {\em relative} fluctuations (which  we
measure) are proportional to $\sqrt{N}/N
= 1/\sqrt{N}$, and therefore increase with decreasing surface brightness.

The measured fluctuations in each of six annuli are
shown in the right panel of Fig.\ \ref{sbf_quant.fig}.
The rms increases with the distance to the center of the galaxy,
from 0.006 in the inner annulus to 0.024 in the outer aperture.
%These values imply that the ``effective'' number of
%giants per (PSF-convolved) pixel
%ranges from $\approx 28,000$ in the central regions
%to $\approx 1700$ near the edge of the field.\footnote{The
%effective number of giants is defined here as the number of stars per
%pixel if all stars had the same luminosity. With that (incorrect)
%assumption, the number is simply 1/SBF$^2$, with SBF the rms surface
%brightness fluctuation.}
The orange
curves in Fig.\ \ref{sbf_quant.fig} show the expected change with
aperture under the assumption that the number of giants is
proportional to the surface brightness of the galaxy. This simple
expectation fits the observations well (see {Tonry} \& {Schneider} 1988).
The small asymmetry in the observed distributions may be partially caused
by faint globular clusters and background galaxies.
For the main analysis in the following sections
we select all $9\times 10^5$ pixels with model surface brightness
$\mu_{\rm F814W}<18.8$\,mag\,arcsec$^{-2}$.

\subsection{Averaged Narrow-band Fluctuations}

In contrast to previous studies we are not concerned with an absolute
calibration of the surface brightness fluctuation signal. 
The goal of this study is to measure the spectral variation as
a function of the surface brightness fluctuation, that is, to determine
by how much the spectrum of a pixel changes for a given change in its
brightness. This relative nature of the method is an important aspect,
as it makes the analysis independent of the distance of
the galaxy, the effects of the PSF, and other effects that determine
the amplitude of the fluctuation signal. As discussed earlier,
the only requirement is
that the PSF and astrometry of the
narrow-band observations are matched to those of the
broad-band observations.

We measure the spectral variation by dividing adjacent narrow-band
filters. Four filter combinations are used: F823N/F836N, F836N/F856N,
F856N/F880N, and F980N/F995N. For each of these combinations an
``index image'' was created by taking the ratio of the two
adjacent narrow-band images.
These index images were normalized by dividing by the median value of
the index. In principle, the images now contain the relative index
variation, with respect to a median index value that is identical
to 1, for each pixel (in addition to noise). However, the median
is not exactly 1 in the entire image, due to relative errors in
flat fielding and a possible contribution from real
large scale spatial variation in the index. As we are interested only
in the relative fluctuations in the indices and not in their
absolute values, we divided the index images by a ``correction
flat'' that
contains this large scale variation.
This correction flat was
created by filtering the index images with
a median filter of $31\times 31$ pixels. The peak-to-peak variation
in the correction flats is $\lesssim 0.1$\,\%.

\begin{figure}[hbtp]
\epsfxsize=8.5cm
\epsffile[25 180 552 672]{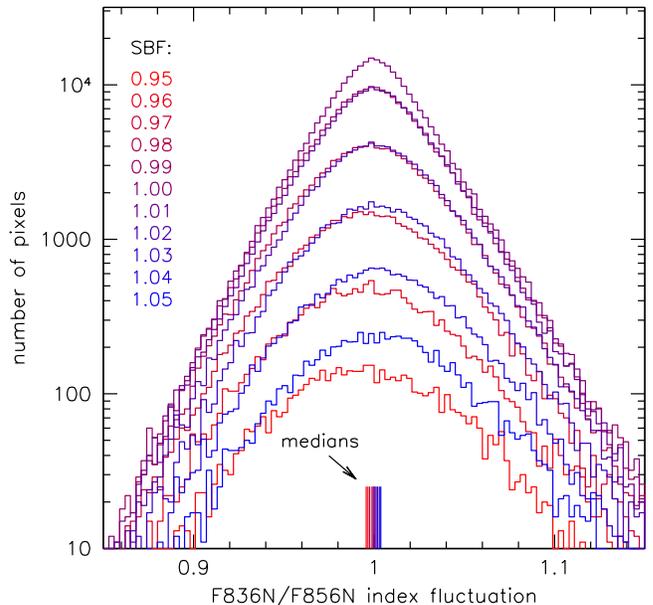}
\caption{\small
Index fluctuation measurements, for the N836N/F856N index. 
The histograms show the measured N836N/F856N index fluctuations
in bins of F814W surface brightness fluctuation. The width of these
histograms is caused by noise: the $1\sigma$ per-pixel uncertainty is
approximately $0.03$. Vertical lines indicate the centers of
the distributions of index
fluctuations; they range from 0.996 to 1.003, for F814W fluctuations in
the range 0.95 -- 1.05.
These centers (estimated with the biweight, which is very close
to a true  median) can be determined with an accuracy of
$\sim 0.0005$; these are the values that are on the vertical axis
of Fig.\ 9.
\label{sbf_raw.fig}}
\end{figure}

The S/N ratio in the
index images is much lower  than in the F814W
images, due to the much smaller width of the ramp filters. The typical
per-pixel
measurement uncertainty in the radial range of interest
ranges from 0.025 -- 0.06, depending on the
index.  As we will
show later, the indices need to be measured with an accuracy of
$\sim 0.0005$ to constrain model parameters -- two orders of magnitude
better than the per-pixel uncertainty.
Fortunately, we can obtain the required S/N ratio
by binning: as we have a precise
measurement of the surface brightness fluctuation
of each pixel from the F814W images, we can average
all pixels in the index images that are known to have a particular fluctuation.
This procedure is demonstrated for the F836N/F856N index
in Fig.\ \ref{sbf_raw.fig}. The rms scatter in individual index
measurements (the $1\sigma$ width of the histograms) is $0.031$.
Nevertheless, in each surface brightness
fluctuation bin the average index
value can be determined with an uncertainty of
only $\sim 0.0002$ (not including systematic errors),
due to the large number of pixels that is averaged.
We experimented with different averaging methods and chose
the biweight ({Beers}, {Flynn}, \& {Gebhardt} 1990), as it is both robust and insensitive
to outliers. Using the median instead leads to nearly indistinguishable
results. We also explored whether the results depend on the radius from
the center of the galaxy, the flat fielding, sky subtraction, and errors
in the shifts. The systematic error due to these effects is estimated
at 0.0005, that is, 0.05\%. 
This uncertainty was added in quadrature to the Poisson error of each
index measurement.

\begin{figure*}[hbtp]
\epsfxsize=15.5cm
\begin{center}
\epsffile[50 195 547 672]{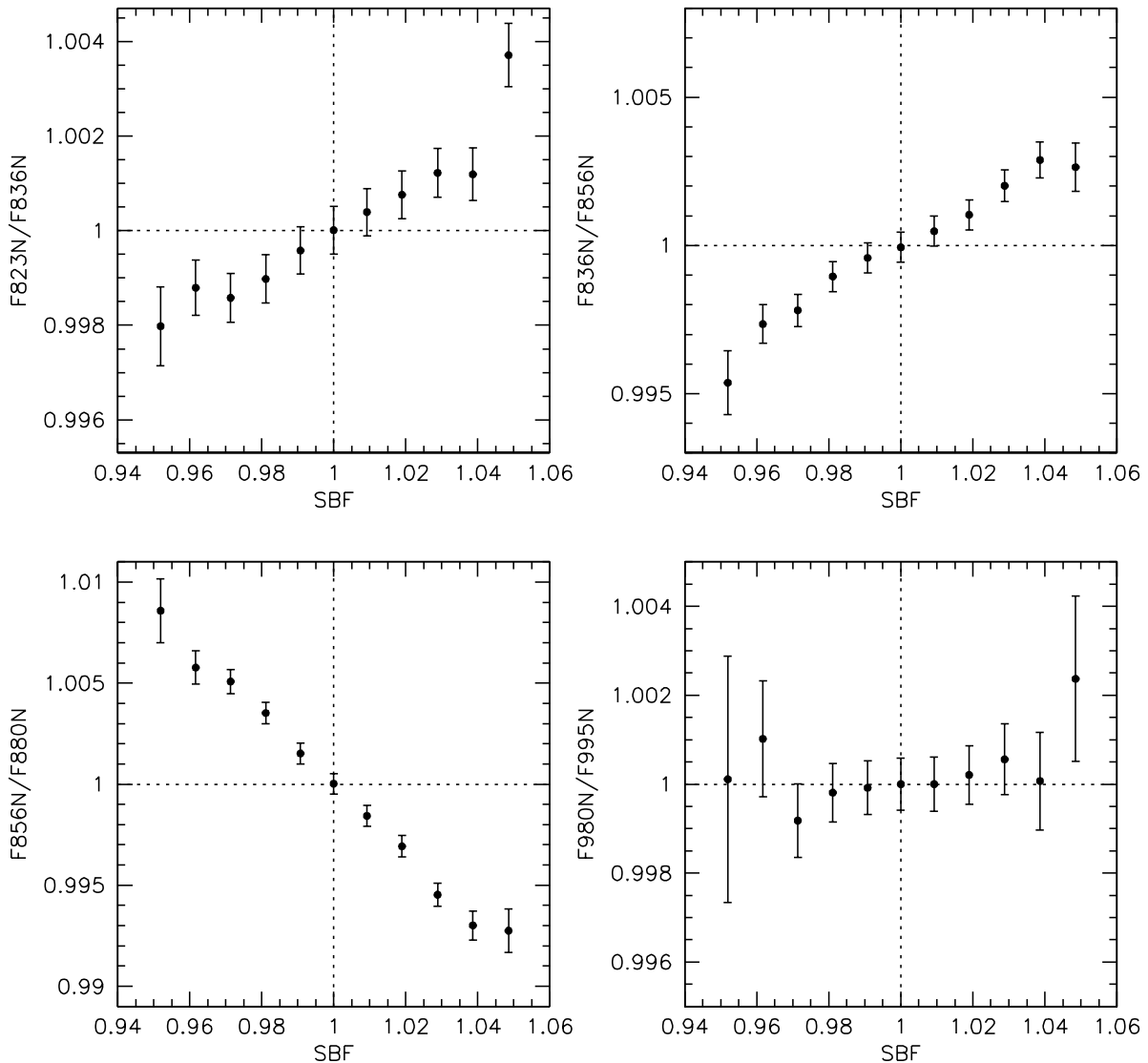}
\end{center}
\caption{\small
Averaged ramp filter index measurements as a function of surface
brightness fluctuation. The four panels show the
filter combinations F823N/F836N,
F836N/F856N, F856N/F880N, and F980N/F995N. 
For three of the filter combinations there is a clear relation between
the value of the index and 
the strength of the surface brightness fluctuation. Errorbars are
a combination of the expected Poisson noise and
a 0.05\% contribution from systematic uncertainties.
The index values are 1 for SBF\,=\,1 by construction.
\label{fluct_meas.fig}}
\end{figure*}

The relations between the
averaged measured F823N/F836N, F836N/F856N, F856N/F880N,
and F980N/F995N indices and the surface brightness
fluctuation are shown in Fig.\ \ref{fluct_meas.fig}.
There are clear relations, but
the variation is small:
over the full range of $\pm 0.05$ fluctuations the indices vary by
maximally $\pm 0.002$
for F823N/F836N and F980N/F995N, $\pm 0.004$ for
F836N/F856N, and $\pm 0.01$ for F856N/F880N. 
We note that the relation between the index strength and the surface brightness
fluctuation is independent of the spatial resolution of the observations,
as long as the quantity on the vertical axis is derived from data with
the exact same spatial resolution as the quantity on the horizontal axis.
The spatial resolution determines the dynamic range of the fluctuations
that are probed at fixed surface brightness, not the form of the
correlations between narrow-band indices and the
surface brightness fluctuations.\footnote{Except in the
regime of very large fluctuations, where the relations are no longer
linear; see \S\,\ref{large.sec}.}

We conclude from Fig.\ \ref{fluct_meas.fig} that pixels with high
fluctuations have systematically different narrow-band indices than
pixels with low fluctuations.
In other words, sightlines through \targ\ with a
relatively high number of giant stars produce a different spectrum than
sightlines with a relatively low number of giant stars.
The fact that two indices show a positive correlation and one
shows a negative correlation implies that we are seeing
variation in spectral {\em features}, not an overall tilt in the
spectrum. In the next Section we interpret the correlations
in Fig.\ \ref{fluct_meas.fig} in the context of expectations
from the stellar spectra that go into the
{Conroy} \& {van Dokkum} (2012a) stellar population synthesis models.

\begin{figure*}[hbtp]
\epsfxsize=16.5cm
\begin{center}
\epsffile[18 200 592 718]{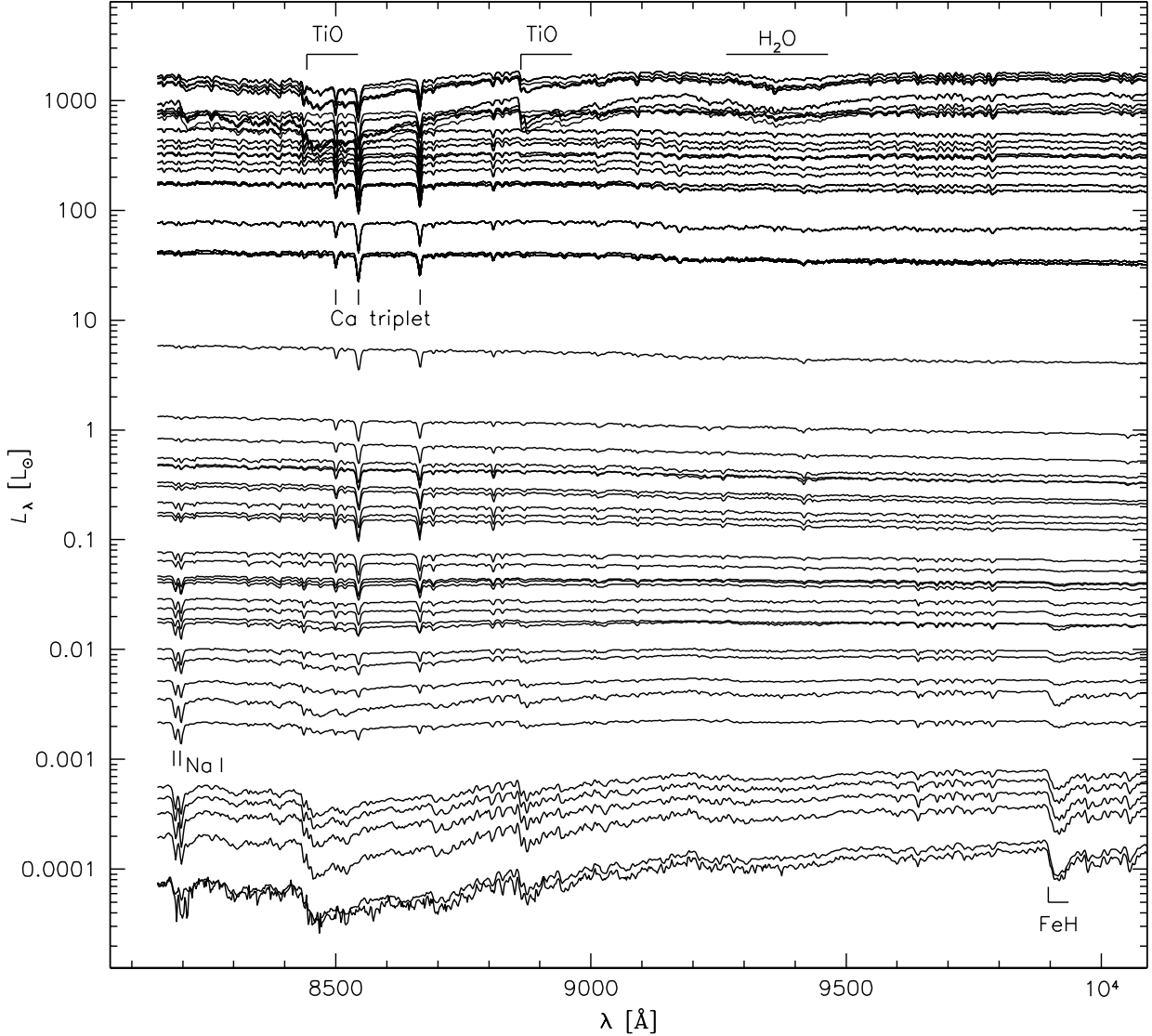}
\end{center}
\caption{\small
Spectra of the stars that are used in the Solar metallicity, 13.5\,Gyr model.
These are the same templates
as used in Conroy \& van Dokkum (2012a); the spectra are from the IRTF
library (Rayner et al.\ 2009).
The most luminous stars have prominent
Ca triplet lines as well as  strong
TiO and H$_2$O molecular
absorption bands. The faintest stars have strong
molecular lines, strong Na\,I
and FeH Wing-Ford absorption, and weak Ca triplet lines.
\label{allstars.fig}}
\end{figure*}

\section{Modeling of Fluctuation Spectra}
\label{model.sec}
The observed
narrow-band index variations are modeled by mimicking the observations
as closely as possible.
We simulate the stellar population in a pixel by Poisson sampling
stars from a stellar library according to an isochrone,
and then calculate the total luminosity and
the integrated spectrum for each of these pixels. Next, we determine the
relations between the narrow-band indices and the
surface brightness fluctuations
in the same way as was done for the data, taking the redshift and velocity
dispersion of the galaxy, the filter curves, and the uncertainties in the
data into account.

\subsection{Model Ingredients}
\label{ingredients.sec}

We use the same empirical stellar spectra that
{Conroy} \& {van Dokkum} (2012a) use in their synthesis
of old stellar populations. The number of templates depends on the
model age; in our default 13.5\,Gyr model 106 templates are used.
They are actually 70 distinct stellar spectra,
with some of them doing ``double duty'' in different evolutionary
phases that bring them to the
same region of the HR diagram. They are
carefully vetted Solar metallicity stars from our own Galaxy; their
near-IR spectra come from the IRTF spectral library of
cool stars ({Rayner} {et~al.} 2009). Details on their selection and
the derivation of their
bolometric luminosities and other parameters are given in {Conroy} \& {van Dokkum} (2012a).

The spectra are shown in Fig.\ \ref{allstars.fig}.
In luminous stars
the Ca\,II triplet lines are strong,\footnote{In the most
luminous stars Ca\,II becomes weaker again as the strong TiO absorption
reduces the pseudo-continuum.} whereas Na\,I and  FeH
({Wing} \& {Ford} 1969) are weak. Low luminosity dwarfs have
weak Ca\,II and prominent Na\,I and FeH absorption. These 
characteristics are behind the widespread application of these three
spectral features
to constrain the stellar mass function from the integrated spectra
of old stellar populations
(see, e.g., {Conroy} \& {van Dokkum} 2012b, and references therein).
As discussed in \S\,\ref{obs.sec} the wavelengths of
the ACS ramp filters were chosen to measure the strengths
of these (redshifted) features.\footnote{Another reason
to focus on these three features is that, for nearly
all stars, one
of them is
the strongest absorption feature in the 0.8\,$\mu$m --
1.0\,$\mu$m wavelength range.}

In the most luminous and the least luminous stars the spectra show
broad bumps and troughs characteristic of molecular transitions.
The most prominent of these are H$_2$O and, particularly,
TiO, which has many bands in this wavelength range ({Valenti}, {Piskunov}, \&  {Johns-Krull} 1998).
Two bandheads are marked in Fig.\ \ref{allstars.fig}.
The strength of the H$_2$O and TiO
bands depends on temperature and surface gravity,
and they are strongest in late M giants at the tip of the giant
branch.

\begin{figure}[hbtp]
\epsfxsize=8.5cm
\begin{center}
\epsffile[22 170 592 718]{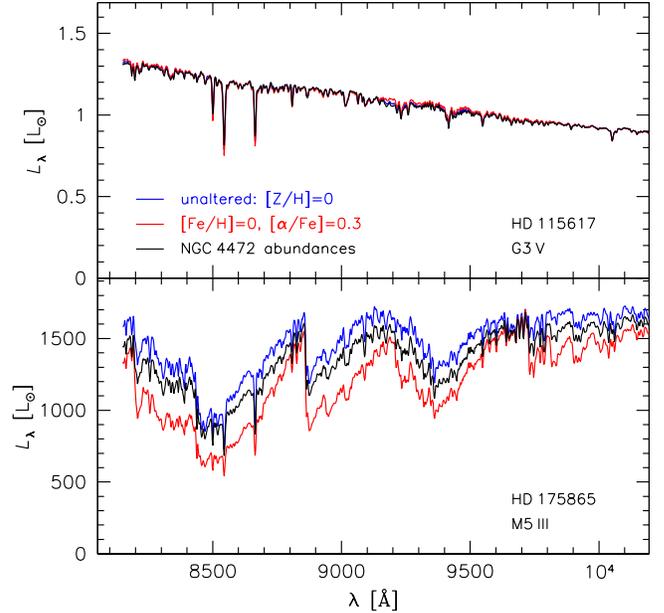}
\end{center}
\caption{\small
Abundance variations are accounted for by adjusting the spectra of all
individual stars in the library. Two example spectra are shown: a Sun-like
star near the main sequence turn-off (top), and a luminous
($L_{\rm bol}\approx 2000L_{\odot}$) red giant (bottom). The original,
approximately Solar metallicity spectra are shown in blue. The red
and black spectra are created by adjusting the blue spectrum
by the theoretical response to changes in the elemental
abundances. The red spectra show the effect of increasing
the abundance of all $\alpha$-elements. The black spectra (used
in our default model)
have an abundance pattern that  is derived from independent
observations of \targ.
\label{Zeffect.fig}}
\end{figure}

The empirical templates were selected to have Solar abundances, but
as is well known elliptical galaxies have non-Solar abundance ratios.
In order to explore the effects of metallicity, and to create
a default model that resembles \targ\ as closely as possible, we
have constructed two stellar libraries with alternative elemental
abundances.
Each individual empirical stellar template
was multiplied by the ratio of two theoretical spectra: one with
the adjusted abundance, and one with Solar abundance. In this
way, the theoretical
response to an abundance change is ``grafted'' onto the empirical
spectra.
This procedure is similar to that followed
in {Conroy} \& {van Dokkum} (2012a); the difference is that in {Conroy} \& {van Dokkum} (2012a)
we adjusted the integrated spectra and here we apply the corrections
to individual stars.
 The first alternative library comprises
``standard'' $\alpha$-enhanced spectra, with [Fe/H]\,=\,0
and $[\alpha/$Fe]\,=\,0.3.\footnote{That is,
[O/Fe]\,=\,[Mg/Fe]\,=\,[Si/Fe]\,=\,[Ca/Fe]\,=\,[Ti/Fe]\,=\,0.3.}
The second, and default, library is
tuned to match the individual elemental abundances of \targ,
as derived from integrated-light long-slit spectroscopy
(Conroy \& van Dokkum 2012a and C.\ Conroy \& P.\
van Dokkum, in preparation). \targ\ has [Fe/H]\,=\,0.05,
[Mg/H]\,=\,0.3, [O/H]\,=\,0.3, [Ti/H]\,=\,0.13, [Na/H]\,=\,0.6,
[Ca/H]\,=\,0.05, [C/H]\,=\,0.1, and [N/H]\,=\,0.35.
In Fig.\ \ref{Zeffect.fig} we show the differences between the spectra
of two stars in the three libraries. For Sun-like stars (top panel) the
differences are subtle, but for luminous late type giants (bottom panel) they
are substantial, with the spectrum that has the NGC\,4472 abundance pattern 
falling in between the (original)
Solar-metallicity spectrum and the $\alpha-$enhanced
spectrum.

\begin{figure}[hbtp]
\epsfxsize=8.5cm
\begin{center}
\epsffile[18 161 592 718]{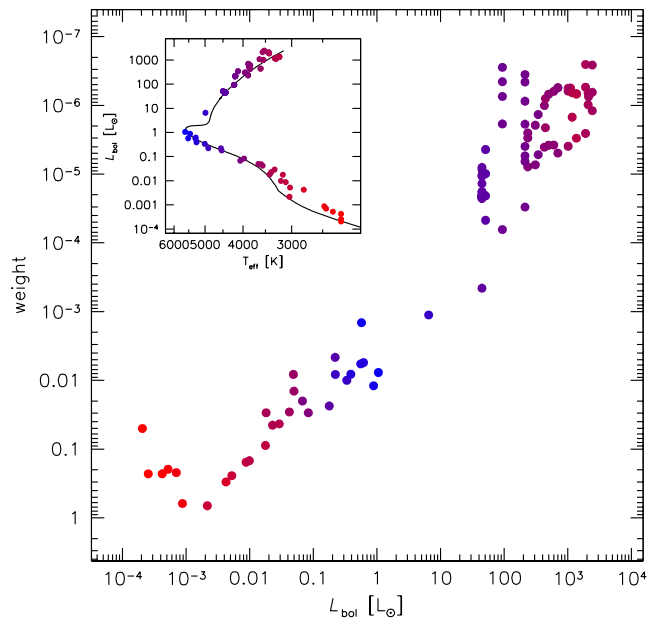}
\end{center}
\caption{\small
Numerical weight of the stars in the default
stellar library, as a function
of their bolometric luminosity. This is for
a Salpeter IMF and an age of 13.5\,Gyr.
%The weights are normalized in such a way that they show
%the expected number of stars of a given type
%in a typical pixel (convolved with the PSF).
The points are color-coded by the effective temperature of
the stars; the inset shows the location of the stars in the HR
diagram. The most luminous giants are rare; there are only a few
dozen stars with $L_{\rm bol}>1000 L_{\odot}$ in a typical
pixel of NGC\,4472.
\label{starweight.fig}}
\end{figure}

Each star in the library is assigned a weight, which is determined by
a combination of a stellar evolution model (as described in
Conroy \& van Dokkum 2012a) and the IMF. As we did for the
metallicity, we base our default model 
on the best fit
to the integrated-light spectrum of \targ\
(see Conroy \& van Dokkum 2012a).
The default weights are therefore
calculated for a 13.5 Gyr old stellar population
with a {Salpeter} (1955) IMF. The weight is defined
such that it is proportional to the
average number of stars in a pixel. The relation between the
weight and the bolometric luminosity is shown in Fig.\ \ref{starweight.fig}.
%In this Figure the weight is normalized such that the total number
%of stars in a PSF-convolved pixel is $2\times 10^7$.
The most luminous giants are rare; only 1 in every 100,000 stars
has $L_{\rm bol}>1000\,L_{\rm \odot}$.

In practice, we simulate $N_{\rm pix}$ ``pixels'' with a per-pixel
mass of $M_{\rm stars}$.
For a Salpeter IMF and an age of 13.5\,Gyr,
the number of stars in each pixel $N_{\rm stars}$
is related to the  mass as $N_{\rm stars}
=3.4 M_{\rm stars}$. We
assign $N_{\rm stars}$ to each pixel.
Each star in each pixel
is labeled with a number from 1 to $n$, with $n$ the
number of templates in the library ($n=106$ for the default library).
This labeling is
based on random numbers, with the probability of assigning a particular
template proportional to its weight.
The total bolometric
luminosity of a pixel is calculated by summing the bolometric luminosities
of its stars. 
We also sum
the stellar luminosities integrated over the
wavelength range 8150\,\AA -- 8400\,\AA,
as a proxy for the total F814W luminosity. 
The integrated spectrum of each pixel is constructed by summing the
spectra of its stars.
Finally, the F814W luminosities
were perturbed with a Gaussian of width $0.006\langle L_{\rm F814W}\rangle$,
to account for measurement uncertainties in the surface brightness
fluctuation measurements (see \S\,\ref{sbfmeas.sec}).

\subsection{Predicted Surface Brightness Fluctuations}

\begin{figure}[hbtp]
\epsfxsize=8.5cm
\begin{center}
\epsffile[25 250 588 718]{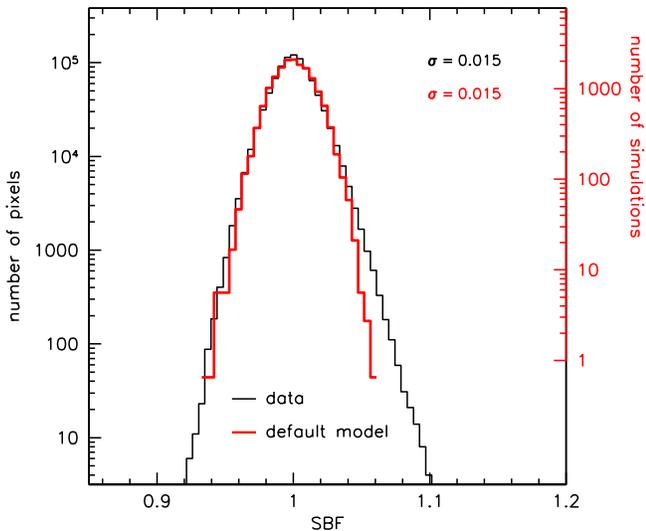}
\end{center}
\caption{\small
Comparison of the modeled F814W surface brightness fluctuations (red)
to the observed ones (black), for $2\times 10^7$ stars per
pixel. This model reproduces
the width of the observed distribution well. The small difference
in the shape of the distribution is caused by the fact that the model
does not take the surface brightness profile of \targ\ into account.
\label{comp_sbf.fig}}
\end{figure}

We first compare the distribution of F814W surface brightness fluctuations
in our default model to the data. In this comparison, the number of stars
per pixel is an adjustable parameter, although we note that our results are not
very sensitive to the precise number of stars that is adopted.
Figure \ref{comp_sbf.fig} compares the observed
surface brightness fluctuations to those in our default model, for
$N_{\rm pix}=2\times 10^4$, $N_{\rm stars} = 2\times 10^7$, and
$M_{\rm stars} = 6\times 10^6$\,\msun.
For this choice
of $N_{\rm stars}$ there is a good match between the distributions, with
an rms in both the data and the observations of 0.015. 
The distributions have a slightly different shape because we did not
take the surface brightness profile of \targ\ into account in
the modeling. 

\begin{figure*}[htbp]
\epsfxsize=16.5cm
\begin{center}
\epsffile[22 442 592 718]{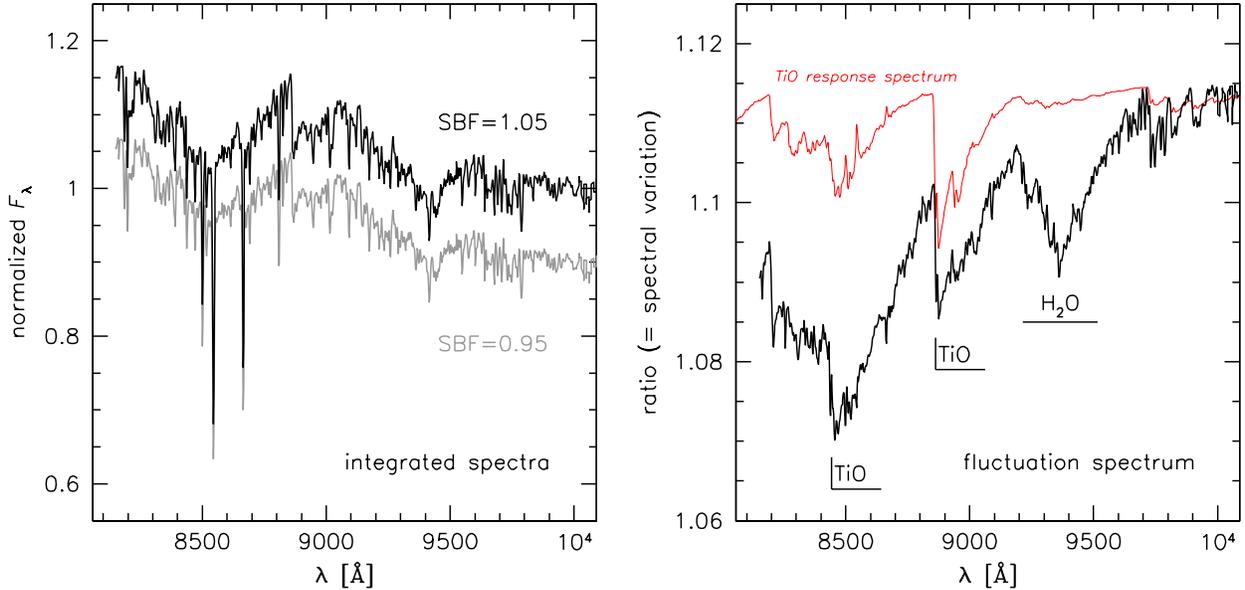}
\end{center}
\caption{\small
Left panel: predicted spectra in the 0.8\,$\mu$m --
1.0\,$\mu$m range of pixels with SBF\,=\,1.05 (black) and SBF\,=\,0.95
(grey), in our default model. Right panel: 
predicted fluctuation
spectrum, obtained by dividing the two spectra shown in the
left panel. The fluctuation spectrum is starkly different from
the integrated spectra, and characterized by molecular absorption bands.
The thin red line shows a TiO response spectrum (see text)
for reference. These TiO
and H$_2$O bands are  prominent in red
giants with $L_{\rm bol}>1000$\,L$_{\odot}$.
\label{fluctspec1.fig}}
\end{figure*}

The amplitude of the surface brightness fluctuations
is particularly sensitive to the number of giants.
As a result, for fixed $N_{\rm stars}$, the model amplitude is a strong
function of the IMF, the age, and other parameters.
In later Sections we investigate the effects of changing the model
parameters, and  we always adjust the number of stars per pixel in order
to match the observed rms of the F814W surface brightness fluctuations.
In this context it should be noted that the parameter $N_{\rm stars}$ is
not the true number of stars in a projected $0\farcs 05
\times 0\farcs 05$ aperture, but  the ``effective'' number of stars per
pixel after smoothing by the ACS PSF (roughly corresponding to
uncorrelated areas of $0\farcs 21 \times 0\farcs 21$). 

We stress that, although we tune the models to reproduce the observed
rms, this is actually not necessary. As explained in the Introduction,
our method is only sensitive to the {\em relative} change in the
spectrum for a given {\em relative} change in the brightness of
a pixel.  Specifically, a model with less stars per
pixel, and therefore a larger absolute fluctuation signal, produces 
(to first order) the
same relations between spectral variation and flux variation as our
default model. The only reason why we adopt a number of stars in
the model that produces the correct fluctuation rms is a subtle
second-order effect:  the smoothing of the modeled F814W
luminosities with the observational error (see \S\,4.1)
can only be done correctly if the absolute rms is matched
reasonably well.

Although it is not important for our analysis, we can ask whether the
observed rms of the fluctuations is similar to the expected rms based
on the distance and surface brightness of \targ\ and the ACS PSF. Phrased
differently, we can ask whether the stellar mass in our model is consistent
with the observed surface brightness of \targ. The median surface brightness
of \targ\ in the analysis region is
$\mu_{\rm F814W}=17.7$\,mag\,arcsec$^{-2}$. For a distance modulus
of $31.17 \pm 0.07$ (Mei et al.\ 2007) this corresponds to
$L_{\rm F814W} = (9.5\pm 2.3) \times 10^5$\,L$_{\odot}$ per $(0\farcs 21 \pm 0\farcs
02)^2$ resolution element.\footnote{The error in the size
of the resolution element reflects
the uncertainty due to the spatial variation in the PSF and the effects of
drizzling.}  The $M/L_{\rm F814W}$ ratio of \targ, as measured from our
Keck LRIS spectrum (Conroy \& van Dokkum 2012b, and in preparation), is
$3.8 \pm 0.5$. The expected stellar mass is therefore $(4 \pm 1)
\times 10^6$\,\msun. The stellar mass in our model is a factor 1.5 higher,
but the difference is barely significant: based on the surface brightness
of \targ\ we would have predicted an rms of $0.012 \pm 0.002$, and
we measure an rms of 0.015.

\subsection{Predicted Fluctuation Spectra}
\label{pred.sec}

We now turn to the spectral variation as a function of the surface brightness
fluctuation. The left panel in Fig.\ \ref{fluctspec1.fig}
shows the average spectra of simulated pixels with fluctuations
of $+5$\,\% and $-5$\,\%.
The spectra are offset by $\sim 10$\,\%, as expected,
but are otherwise very similar: they are both dominated by
the well-known spectral features of old stellar
populations (see, e.g., the top panel of Fig.\ \ref{filters.fig}).

The right panel shows the ratio of the two spectra, the
``fluctuation spectrum''. The fluctuation
spectrum is very different from the
integrated-light spectra. There are
no obvious features at the locations of the Na\,I $\lambda 8183,8195$
doublet, the Ca II triplet, or the FeH $\lambda 9916$ Wing-Ford
band. Instead, the fluctuation spectrum is characterized by sharp
bandheads at $\approx 8440$\,\AA\ and $\approx 8870$\,\AA, as well
as a broad feature at $\approx 9350$\,\AA. We identify these features
as TiO bandheads and H$_2$O absorption, respectively. This is
demonstrated by the thin red line, which
is an approximation of a ``pure'' TiO spectrum.
This TiO spectrum was generated by turning TiO on and off in the
modeling of the atmosphere of a single star with $T_{\rm eff}=3750$\,K
and $\log(g)=1.0$ and then taking the ratio of the resulting
spectra.

As can be seen in Fig.\ \ref{allstars.fig} 
the TiO and H$_2$O bands are strong in the most luminous giants, which have
$L_{\rm bol}>1000$\,L$_{\odot}$ and $T_{\rm eff} = 3200$\,K -- 3600\,K.
In the 
regime of small fluctuations, a pixel's
surface brightness fluctuation
is essentially determined by the number of these cool,
luminous giants it contains. In our default model with
$N_{\rm stars} = 2\times 10^7$ stars per pixel, the average number
of stars with $L_{\rm bol}>1000$\,L$_{\odot}$ per (PSF-convolved)
pixel ranges from 91
for pixels with
SBF\,=\,0.95 to 150 for pixels with SBF\,=\,1.05.

\subsection{Comparison to Observations}

We now compare the predicted fluctuation spectra to the observed
relations between the narrow-band indices and the F814W surface
brightness fluctuations.
We averaged the model spectra in each of 11 surface brightness
fluctuation bins, from 0.95 to 1.05 with steps of 0.01.
The spectra  were redshifted to the distance
of \targ\ (997\,\kms) and smoothed to a resolution
of $\sigma = 300$\,\kms, which is the
approximate velocity dispersion of \targ. Then, the
spectra were divided by the spectrum of the central
surface brightness fluctuation bin, with SBF\,=\,1.
This gives 11 differential fluctuation spectra, appropriate for fluctuations
ranging from $-0.05$ to $+0.05$. The central fluctuation spectrum is unity
at all wavelengths, by construction.

\begin{figure}[hbtp]
\epsfxsize=8.5cm
\epsffile[25 165 592 718]{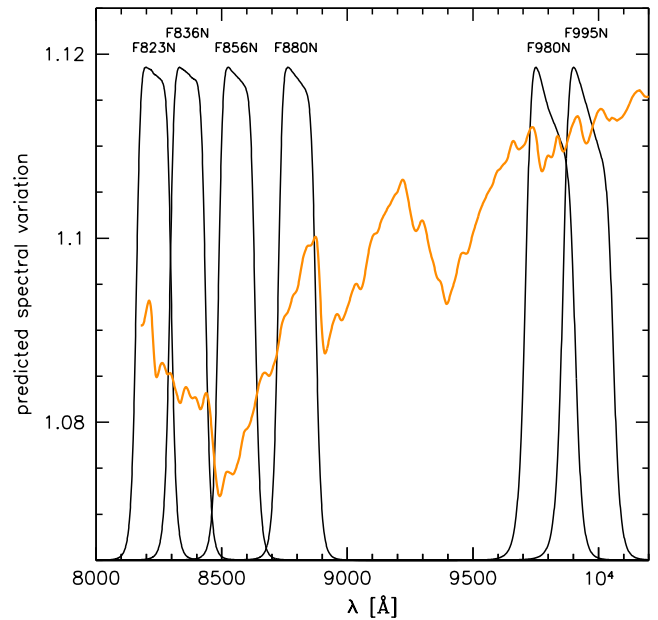}
\caption{\small
Comparison of the ACS ramp filter locations to the predicted spectral
fluctuations in our default model. The model was redshifted to the
distance of \targ\ and smoothed to a velocity dispersion of
 300\,\kms. The four bluest ramp filters sample the strong spectral
variation expected from the TiO features in the most luminous giants.
\label{smooth.fig}}
\end{figure}

In Fig.\ \ref{smooth.fig} we show the ratio of the two extreme
spectra, those with SBF\,=\,1.05 and SBF\,=\,0.95.
This spectrum can be compared to the unredshifted, unsmoothed
version shown in Fig.\ \ref{fluctspec1.fig}.
The six ramp filters are overplotted, after normalizing the
filter curves by their peaks.
The four blue filters 
sample wavelengths where there should be
significant variation due to TiO, with the two reddest filters
effectively acting as a control.
The 11 fluctuation
spectra were integrated using these synthetic ramp filter
curves. Finally, the ratios between adjacent ramp filter fluxes
were determined, creating synthetic indices that can be compared
to the data.

\begin{figure*}[hbtp]
\epsfxsize=18cm
\begin{center}
\epsffile[18 430 592 718]{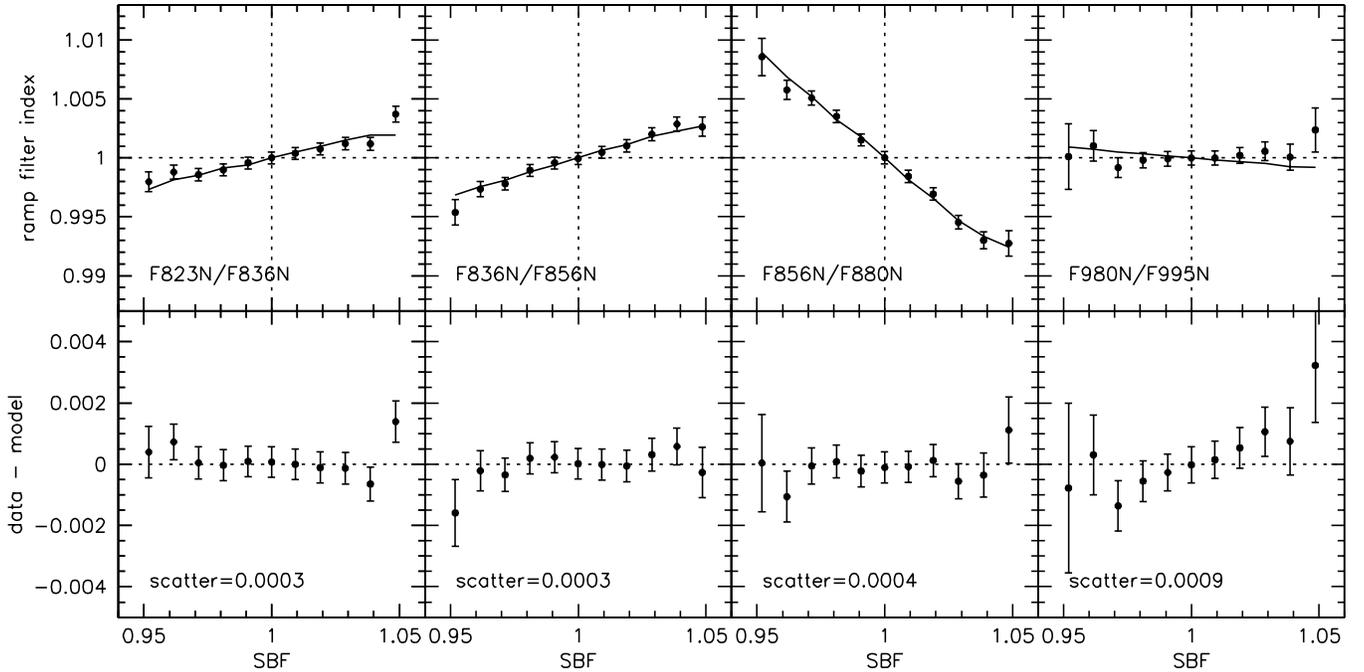}
\end{center}
\caption{\small
Comparison of observed ramp filter index variations to predictions from
our default model. In the top panels, the points are identical to those
in Fig.\ \ref{fluct_meas.fig}, and the lines are model predictions.
The agreement is excellent, which means that we have
a good understanding of both the
origin of the surface brightness fluctuations and of the spectra of
the most luminous giants in massive early-type galaxies.
The bottom panels show the residuals from the predicted correlations.
The scatter in the residuals is $<0.1$\,\% in all cases, and
lower than the systematic uncertainty of $0.05$\,\% in three of the
four panels.
Note that there are no free parameters
in the model: the stellar library is that of Conroy
\& van Dokkum (2012a), and the abundance pattern and age come from
a fit to the integrated-light spectrum of \targ.
\label{moddata1.fig}}
\end{figure*}

In Fig.\ \ref{moddata1.fig} the model predictions are compared to the
observed fluctuations in the ramp filters. The data points are the same
as those in Fig.\ \ref{fluct_meas.fig}; the only difference is that
the vertical axis spans the same range in all panels of Fig.\ \ref{moddata1.fig}.
The lines are predictions from our default model, that is, a
model that is based on the stellar library and stellar evolution
tracks discussed in Conroy \& van Dokkum (2012a), with an age
of 13.5\,Gyr, the \targ\ abundance pattern as measured from its
integrated-light spectrum, a {Salpeter} (1955) IMF, and a
stellar mass of $6\times 10^6$\,\msun\
per pixel.

The models provide a remarkably good description of the data, even though
they are determined in a completely independent way. Both the sign
of the correlations and their amplitude are a close match. The bottom
panels of Fig.\ \ref{moddata1.fig} show the residuals, obtained by
subtracting the model predictions from the data. The $1\sigma$
scatter in the residuals is $<0.0005$ for three of the
indices. It is slightly higher at $0.0009$ for the reddest index,
and there is some evidence for a systematic trend in the residuals
for this index.
Taking all four indices together,
the reduced $\chi^2$ is 0.6, which suggests that the systematic
error (0.0005) is slightly overestimated. For a systematic error
of 0.00025 the reduced $\chi^2$ is close to 1.

The excellent correspondence between the observed relations
between the ramp filter indices and the F814W surface brightness
fluctuations has several implications: it means
that we understand the origin of the surface brightness fluctuations
in this regime,
and also that our stellar population synthesis of the upper RGB
and the AGB
is correct to within the uncertainties of the method.
%Specifically, this is the first test of the accuracy of
%the modeling of the rare, luminous cool giants that account for
%$\sim 20$\,\% of the integrated luminosity of early-type galaxies.

\begin{figure*}[hbtp]
\epsfxsize=17.5cm
\begin{center}
\epsffile[30 300 592 455]{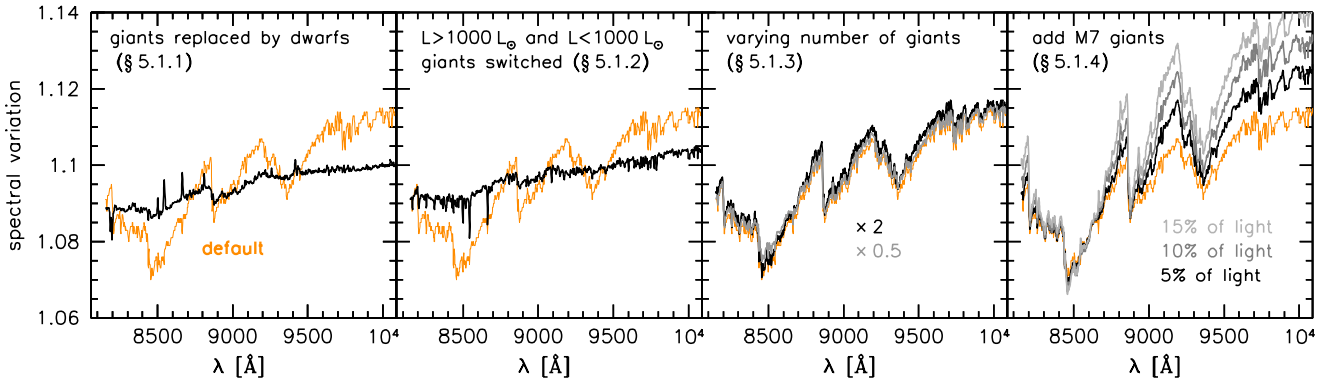}
\end{center}
\caption{\small
Predicted fluctuation spectra when the underlying stellar population
synthesis ingredients
are changed.
Each panel shows the ratio of the average spectra of pixels with high
(SBF\,=\,1.05) and low (SBF\,=\,0.95)
fluctuations. The default model (\S\,4)
is shown in orange. From left to right: model where all giant
spectra are replaced with the spectra of main sequence stars of
the same effective temperature (\S\,5.1.1);
model where the spectra of the most
luminous giants are switched with those of less luminous giants
(\S\,5.1.2); effect of changing the number of luminous
($L_{\rm bol}>1000$\,L$_{\odot}$) stars by a factor of 2 (\S\,5.1.3);
and the effect of adding a population of very cool giants,
contributing 5\,\% -- 15\,\% of the total luminosity.
\label{fluctspec2.fig}}
\end{figure*}

\section{Effects of Varying Model Parameters}

In the previous section we have shown that
our default model provides an excellent fit to the observed fluctuations.
Here we study the effects of changing various aspects of the model.
We consider three distinct types of variation: changes to the underlying
model (specifically, the relative weights of stars in the stellar
library) (\S\,\ref{library.sec}); varying the
age, metallicity, and the IMF (\S\,\ref{age.sec}); and the regime of
very large surface brightness fluctuations (\S\,\ref{large.sec}).

\subsection{Changes to the Underlying Model}
\label{library.sec}

We investigate four variations of the default underlying
model. Each model
in this section has an age of 13.5\,Gyr, the \targ\ elemental abundance
ratios, and a Salpeter IMF.  The variations
either change the spectra in the library of 106 templates or change the
weights of these templates. Changing the weights effectively means changing
the time that stars spend in that particular phase of evolution.
The first two of the variations are dramatic (and effectively ruled
out by many other observations of stellar populations), and are intended
to illustrate the
qualitative behavior of the fluctuation spectrum. The last two are more
subtle (and more reasonable).

\subsubsection{Replacing Giant Spectra by Dwarf Spectra}

The first model variation we consider is to ignore the known differences
between the spectra of dwarfs and giants and assume that the spectrum
of a star depends only on its temperature. We replace the spectrum
of every giant in the library by that of the dwarf star whose temperature
is closest to that of the giant. As the dwarfs cover
a broader range of temperatures
than the giants (see the inset of Fig.\ \ref{starweight.fig}), every
giant in the library has a close temperature match on the main sequence.
We scale each of the dwarf
spectra in such a way
that the average brightness in the $0.8\,\mu$m -- $1.0\,\mu$m
wavelength range is the same as that of the giant it replaces.

The resulting fluctuation spectrum, again obtained by dividing the
averaged spectrum
of +5\,\% F814W fluctuations by the averaged
spectrum of $-5$\,\% fluctuations, is shown in the left panel
of Fig.\ \ref{fluctspec2.fig}.
The fluctuation spectrum is very different from that of
the default model (shown
in orange). Cool dwarfs generally have weaker molecular features than
cool giants, and we therefore do not see the strong TiO and
H$_2$O bands. Instead, we see that high fluctuations have enhanced
Na\,I and FeH absorption and suppressed Ca\,II triplet lines, as
expected from the spectra of cool dwarfs. 

The fluctuation spectra for all 11 surface brightness fluctuation
bins were redshifted and smoothed by the velocity dispersion of
\targ, just as was done for the default model. We calculated
synthetic fluxes in the ramp filters and constructed model indices.
In the top panels of Fig.\ \ref{moddata2.fig} we compared the predictions
from this model to the data, and to the default model. The model
can clearly be ruled out by the data: the spectra
of giants are different from those of
dwarfs of the same temperature.
%The failure of this model, combined
%with the success of the default model, also
%demonstrates spectroscopically that variations in the number of
%giants are the origin of the surface brightness fluctuations.\footnote{It
%was already established beyond reasonable doubt that surface brightness
%fluctuations are caused by variations in the number of luminous giants,
%of course. Nevertheless, the
%observed correlations between the ramp filter indices and the broad band
%fluctuations constitute
%a rather definitive proof.}

\subsubsection{Switching Giants}

With this modification we highlight the fact that the most luminous
giants, with $L_{\rm bol}>1000$\,L$_{\odot}$, are responsible for the
trends in Fig.\ \ref{fluct_meas.fig}.  We switched the spectra of
stars with $L_{\rm bol}>1000$\,L$_{\odot}$ with those of stars with
$L_{\rm bol}=250$\,L$_{\odot} - 1000$\,L$_{\odot}$. No changes were made to
stars with $L_{\rm bol}<250$\,L$_{\odot}$. The rank order within each
set of giants was conserved; that is,
the brightest star with $L_{\rm bol}>1000$\,L$_{\odot}$ was switched with
the brightest star with $L_{\rm bol}<1000$\,L$_{\odot}$.
As before, we scaled each stellar spectrum so it has
the same brightness in the $0.8\,\mu$m -- $1.0\,\mu$m range
as the star it replaces.

The fluctuation spectrum is shown in the second panel of
Fig.\ \ref{fluctspec2.fig}. Again, the broad molecular features have
disappeared, as they now artificially occur in stars that are
much more common and therefore do not contribute significantly
to the surface brightness fluctuation signal. We now see that
the Ca\,II
triplet lines are stronger in pixels with a higher fluctuation.
The reason for this behavior is that these lines
are very strong in stars with
$L_{\rm bol}=250$\,L$_{\odot} - 1000$\,L$_{\odot}$ (see Fig.\ \ref{allstars.fig}).
We will return to this in \S\,\ref{large.sec}, where we discuss
the behavior of the default model in the regime of large, negative
fluctuations (that is, SBF\,$\ll 1$).

In the second row of Fig.\ \ref{moddata2.fig} the observed ramp filter
indices are compared to the predictions from this model. Again, the
model is a poor fit: the predicted trends are too weak, due to the
absence of strong molecular bands in the fluctuation spectra.
As noted earlier, stars with $L>1000$\,L$_{\odot}$ are rare but contribute
significantly to the integrated light.  Their
treatment is an important source of uncertainty in stellar population
synthesis models, and our results provide strong confirmation of the
quality of their implementation.

\subsubsection{Varying the Contribution of Luminous Giants}

Next, we ask what the effect is of changing the contribution of 
stars with $L_{\rm bol}>1000$\,L$_{\odot}$ to the total
luminosity. This contribution depends
on the product of the time spent in their particular phase of
evolution and their luminosity. The luminosities are generally
better constrained than their timescales, and so we vary the
weights 
of the stars (i.e., how many there are
in a pixel) rather than their individual luminosities.

\begin{figure}[hbtp]
\epsfxsize=8.5cm
\epsffile[27 165 592 718]{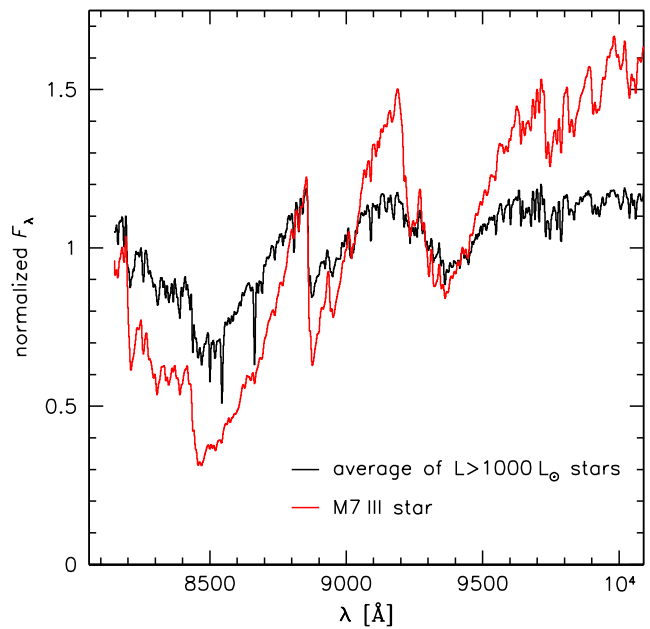}
\caption{\small
Spectrum of the M7 giant HD\,108849, compared to the average spectrum
of stars with $L_{\rm bol}>1000$\,L$_{\odot}$ in our default library.
These stars have spectral types M2 -- M6.
The M7 star is completely dominated by molecular bands. It has
TiO features at the locations of both the
Na\,I $\lambda 8183,8195$ doublet and the
Wing-Ford FeH $\lambda{}9916$ band.
\label{coolgiant.fig}}
\end{figure}

The third panel of Fig.\ \ref{fluctspec2.fig} shows the
effect of increasing (black) and decreasing (grey) the number of
luminous giants by a factor of 2. There is an effect on the fluctuation
spectrum, with the molecular features becoming stronger for an
increasing number of luminous giants, but the  effects are small.
The reason for this modest change is that the spectral change
is, in essence, a second-order
effect. To first order, the variation in both the integrated
luminosity of pixels
and in the normalization of their spectra increases when the number
of giants increases. The variation in the spectra
at fixed luminosity variation is a more subtle effect.

\begin{figure*}[htbp]
\epsfxsize=16cm
\begin{center}
\epsffile[25 167 592 718]{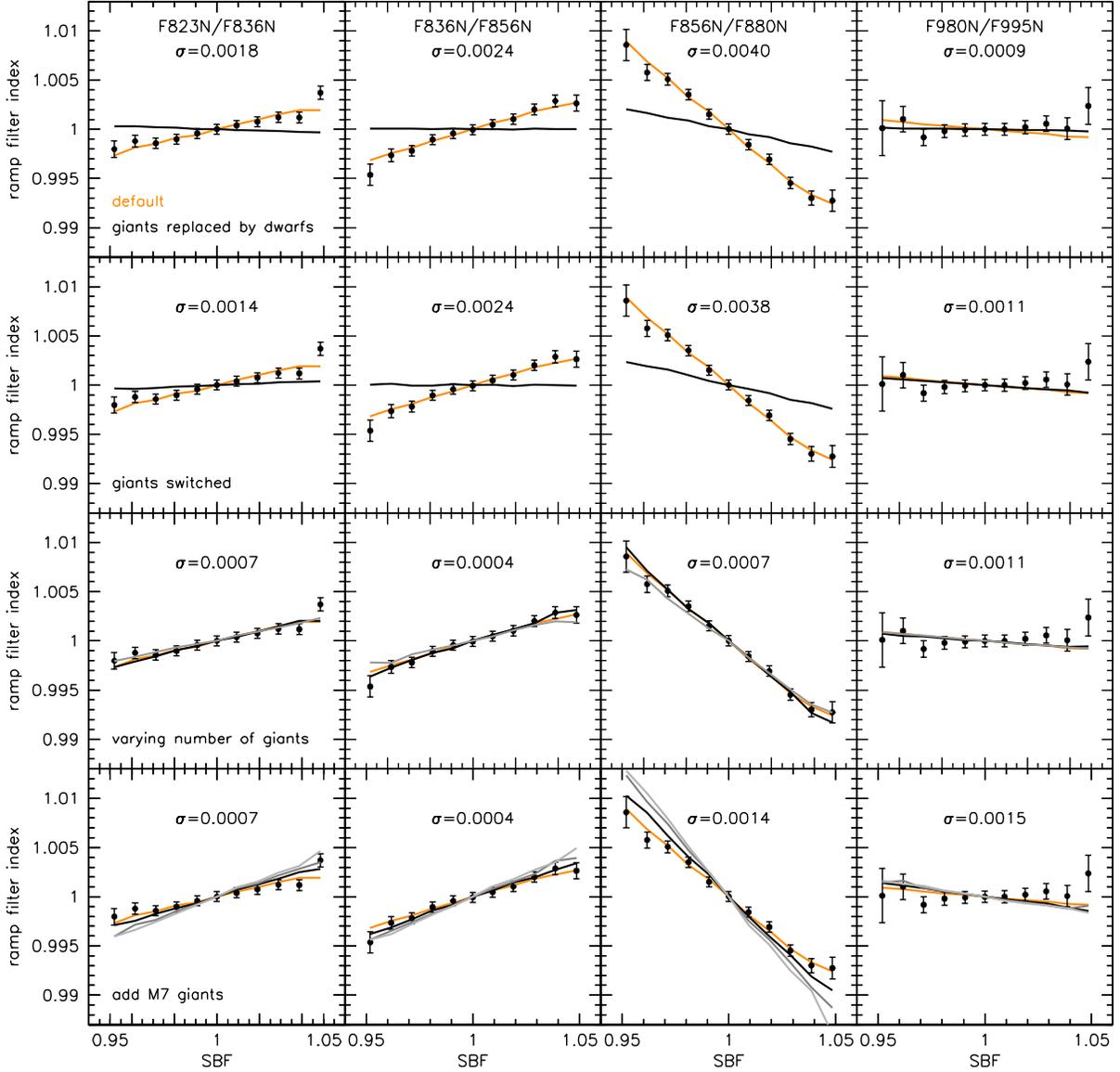}
\end{center}
\caption{\small
Comparison of the observations to the four model variations discussed
in \S\,\ref{library.sec}.
Dramatic changes, such as replacing all giant spectra with those of
main sequence stars of the same temperature, are very poor
fits and can be readily excluded. There is little constraint on the
number of giant stars (third row): a factor of 2 change, in either
direction, produces nearly
identical fits to the data. However,
we can rule out a significant additional contribution of very cool giants
to the light (bottom row). The lines correspond to contributions
of 5\,\%, 10\,\%, and
15\,\% to the light. Even a 5\,\% contribution to the light produces
trends that are too steep, with $\sigma = 0.0014$ in the F856N/F880N
index.
\label{moddata2.fig}}
\end{figure*}

This is also evident in the third row of Fig.\ \ref{moddata2.fig}, where
the data are compared to the model predictions. The scatter in the
residuals is $<0.001$, and similar to the scatter in the residuals
from the default model. The reduced $\chi^2$ is 0.6, whereas it is 10.7
and 9.4 for the models considered in \S\,5.1.1 and \S\,5.1.2 respectively.
To constrain the absolute number of giants an accurate calibration of the
absolute surface brightness fluctuation signal is required.

\subsubsection{Adding Cool Giant Light}

Finally, we consider the possibility that massive elliptical galaxies
have a population of very cool giants that is not included in
standard stellar population synthesis models at old ages. 
The strong TiO bands in these stars overlap with several IMF-sensitive
features, which is why an arbitrary contribution of M giants is included
as one of the
nuisance parameters in the {Conroy} \& {van Dokkum} (2012b) model.
The spectrum of such a cool giant is shown in Fig.\ \ref{coolgiant.fig}.
This M7\,III star, HD\,108849, has a luminosity
$L_{\rm bol} = 2150$\,L$_{\odot}$ and an effective
temperature $T_{\rm eff} = 3075$\,K.
It has much stronger molecular bands than the average spectrum
of $L>1000$\,L$_{\odot}$ stars in the default library (which have
spectral types M2 -- M6).

In the fourth panel of Fig.\ \ref{fluctspec2.fig} we show the
fluctuation spectrum for a model where we add a population of these
cool giants to the light. The weight of this new template
is determined by setting
the contribution of these stars to the total integrated light in the
$0.8\,\mu$m -- $1.0\,\mu$m range to 5\,\%, 10\,\%, and 15\,\%
(black, dark grey, and light grey in Fig.\ \ref{fluctspec2.fig}).
As might have been expected
the fluctuation spectrum shows stronger molecular features than
the default model.

\begin{figure*}[hbtp]
\epsfxsize=20cm
\begin{center}
\epsffile[-5 305 560 455]{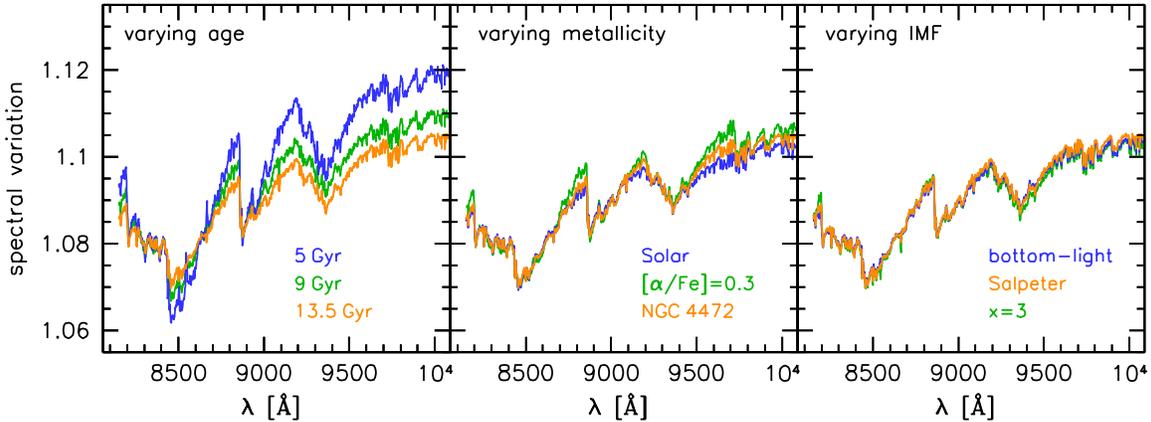}
\end{center}
\caption{\small
Predicted fluctuation spectra for different ages, metallicities, and
IMFs. 
Each panel shows the ratio of the average spectra of pixels with
positive and negative 5\,\% fluctuations (that is, SBF\,=\,1.05
and SBF\,=\,0.95). The predicted spectra depend
on the age and metallicity, but not on the IMF. The
default model, shown in orange in all panels, is based on the
best fit to the integrated-light spectrum of \targ.
\label{fluctspec3.fig}}
\end{figure*}

As shown in the bottom row of Fig.\ \ref{moddata2.fig}, even the
model with the smallest contribution from M7 giants is clearly
a worse fit to the data than the default model. The scatter in
the residuals from the fit is 0.0014 for the F856N/F880N index,
compared to 0.0004 in the default model. The reduced $\chi^2$ is
1.5 for the 5\,\% model (with 43 degrees of freedom),
3.4 for the 10\,\% model, and 6.0 for
the 15\,\% model. The 5\,\% model can be rejected with 98\,\%
confidence, and the other models with $>99$\,\% confidence.

\subsubsection{Summary of Results}
From the four modifications that we considered we can infer that the
data are sensitive to the spectra of the most luminous giants
(those with $L\gtrsim 1000$\,L$_{\odot}$). This conclusion
applies to the regime of small fluctuations (with an rms of 0.01 -- 0.02),
and the wavelength range 0.8\,$\mu$m -- 1.0\,$\mu$m.
The data are remarkably {\em in}sensitive
to the {\em number} of these stars, as shown in the third row of Fig.\
\ref{moddata2.fig}. As explained in \S\,\ref{pred.sec}
this is because varying
the number of these stars 
``stretches'' the relations between narrow-band indices and
the surface brightness fluctuations, but does not change their slope.

Of the four modifications, the first two (\S\,5.1.1 and \S\,5.1.2) mostly
serve to illustrate the fact that the fluctuations are only sensitive to
the most luminous giants. The latter modifications (\S\,5.1.3 and
\S\,5.1.4) are non-trivial and ``realistic'', as the
lifetimes and temperatures of luminous AGB stars are not well-known
from first principles.
Perhaps the most directly applicable result is that
we can exclude
a significant contribution from giants that are cooler
than the lowest temperature ones in our default $t=13.5$\,Gyr
library.
The formal upper limit on their contribution is 3\,\% (with
95\,\% confidence), although we note that this ignores unknown
systematic effects as well as
possible degeneracies with other parameters. This constraint can be
used in stellar population synthesis models to limit the
allowed parameter space (see, e.g., Supp.\ Fig.\ 2 in van Dokkum \&
Conroy 2010 and the discussion in Conroy \& van Dokkum 2012b).

\subsection{Varying Age, Metallicity, and the IMF}
\label{age.sec}

In this Section we fix the
underlying components of the model
and consider variation in the stellar population
parameters age, metallicity, and IMF. In our default model
these parameters are taken from the stellar population synthesis
model that provides the best fit to the high quality Keck integrated-light
spectrum of \targ\ (Conroy \& van Dokkum 2012a; C.\ Conroy \& P.\ van
Dokkum, in preparation). This fit has an age of 13.5\,Gyr,
an IMF that is close to Salpeter, and
the elemental abundance pattern that is described in \S\,\ref{ingredients.sec}.
As we showed in the previous section, the
ramp filter data samples molecular bands in
stars on the upper giant branch; even though we
use the same underlying model as Conroy \& van Dokkum (2012a) it is by
no means
guaranteed that the best-fitting model to the integrated light spectrum
also provides the best fit to the surface brightness fluctuation signal.

\begin{figure*}[htbp]
\epsfxsize=16cm
\begin{center}
\epsffile[25 300 592 718]{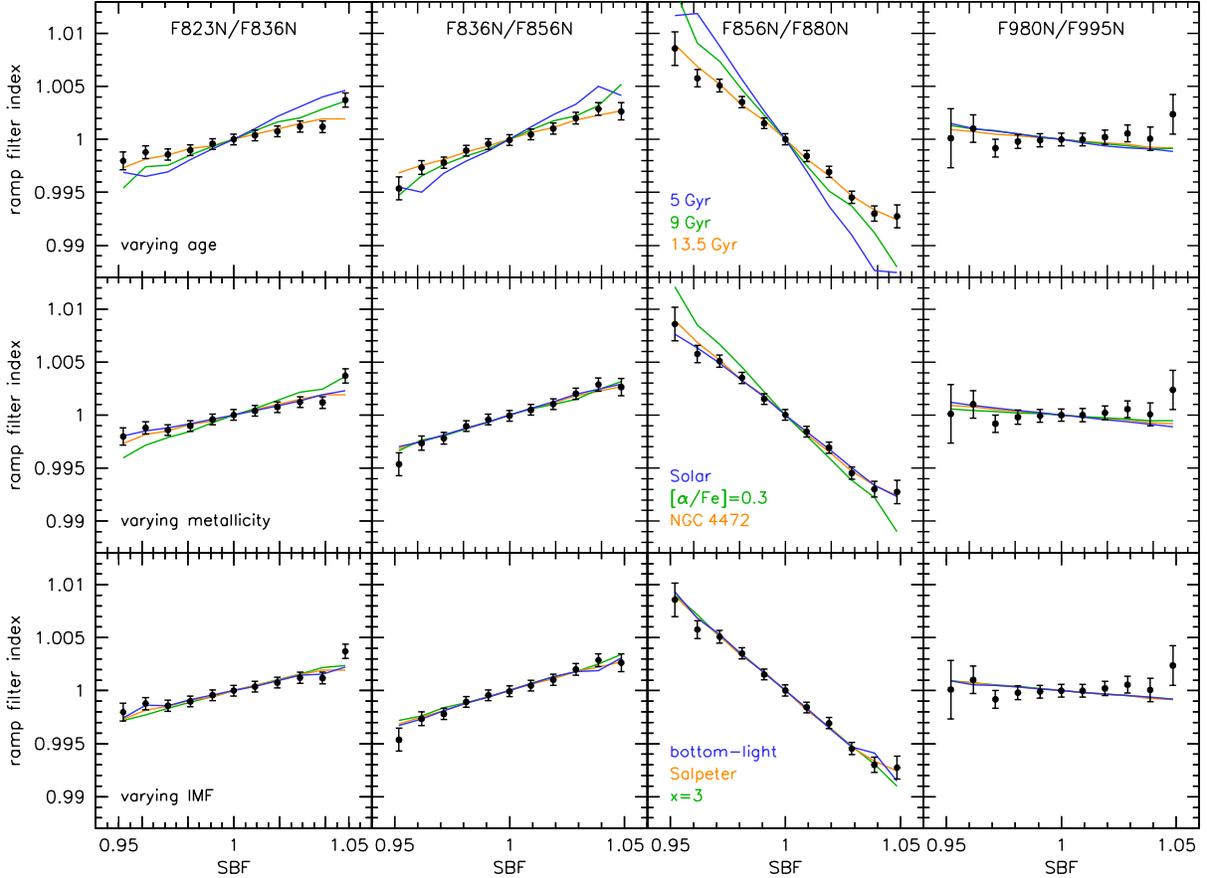}
\end{center}
\caption{\small
Comparison of the observations to models with varying
age, metallicity, and IMF. The ramp filter data are poorly fit with models
that
are younger than $\sim 11$\,Gyr or have a ``standard''
$\alpha-$enhancement of 0.3. The data provide no constraint on the IMF.
Overall the best fits are obtained
for the default model (orange in all plots), which was derived from
integrated-light spectroscopy.
\label{moddata3.fig}}
\end{figure*}

In addition to the default model we consider two younger ages
(9\,Gyr and 5\,Gyr), two alternative abundance patters (Solar and
the ``standard'' $\alpha-$enhanced model described in \S\,4.1),
and two alternative IMFs (a very bottom-heavy IMF with $x=3$, and
a bottom-light {van Dokkum} (2008) IMF with $m_c=2$\,\msun).
The predicted fluctuation spectra are shown in
Fig.\ \ref{fluctspec3.fig}, and the predicted ramp indices are compared
to the observations in Fig.\ \ref{moddata3.fig}.

We find that the indices are quite sensitive to age. The 9\,Gyr model
can formally be excluded at $>99$\,\% confidence.
The reason for this sensitivity is that younger
ages have a greater contribution of late M giants. In particular, the number
of M6 giants is a factor of $\sim 3$ higher in the 9\,Gyr model than
in the 13.5\,Gyr model. As these late type giants -- whose spectra
have very strong molecular absorption bands -- dominate the surface brightness
signal, the change in their contribution at old ages produces a readily
detectable signature. The age constraint from the surface brightness
fluctuations is fully consistent with the
very old age derived from NGC\,4472's integrated-light spectrum, which is
mostly sensitive to the properties of stars near the main sequence
turn-off.  A caveat here is that the measurements are not truly
independent, for two reasons: we are using the same model ingredients
for both constraints, and late M giants would not contribute solely to the
surface brightness signal but also to the integrated light.

The indices are also sensitive to metal abundance, but not in 
a straightforward way.
The standard $\alpha-$enhanced model predicts steeper
relations than observed, and can be ruled out with high
confidence.\footnote{A caveat is that we do not vary the isochrone in
this analysis. When fitting the integrated light spectra we do allow
isochrone variation, as we fit for 
an offset in the location of the main sequence turn-off.
For NGC\,4472, this offset turns out to be small ($\Delta T \sim 10$\,K).}
However, the predicted trends for Solar metallicity are very similar
to those of the default model (and a good fit to the data),
despite the fact that the default
model has Mg and O abundances that are much closer to the $\alpha-$enhanced
model than to the Solar model.
As discussed in
\S\,\ref{ingredients.sec}, the only features that really matter are the
TiO bands, and the Ti abundance of \targ\ is relatively low.\footnote{The
interpretation is complicated by the fact
that the fluctuation spectra effectively measure the varying response of
lines to changes in $T_{\rm eff}$. If lines from
stars of different temperature are on the
same part of the curve of growth,
abundance effects can cancel.}
%Although we
%did not explore this further, we suspect that the Ti abundance (rather than
%the O abundance) drives the strength of the TiO features. If this is correct,
%then our observations constrain [Ti/H] but not any other
%elemental abundances.

\begin{figure*}[hbtp]
\epsfxsize=16.5cm
\epsffile[0 300 592 718]{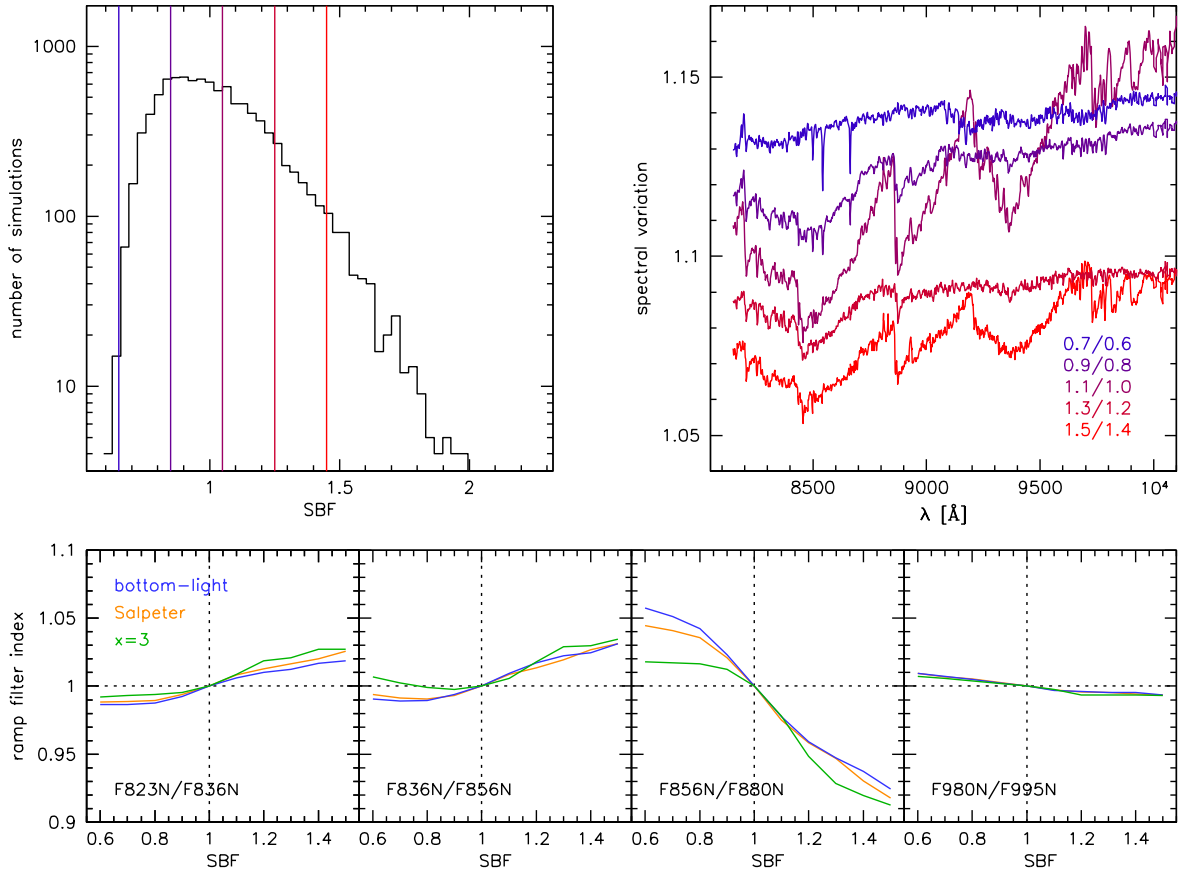}
\caption{\small
The regime of large fluctuations. Top left: predicted distribution
of surface brightness fluctuations for a Salpeter
IMF and $10^5$ stars per pixel, a factor of 200 lower than in our
HST observations of \targ. Top right: fluctuation spectra in different
regimes, from  fluctuations $\ll 1$ to fluctuations $\gg 1$. Large positive
fluctuations are dominated by luminous RGB and AGB stars.
Large negative
fluctuations (SBF\,$<0.9$) probe lower luminosity RGB stars, subgiants,
and main sequence stars.
Bottom panels: ramp filter index predictions for different 
IMFs. There is an IMF dependence, particularly for IMFs steeper than
the Salpeter form.
\label{largefluc.fig}}
\end{figure*}

The data provide no direct constraint on the IMF. The amplitude of
the broad-band fluctuations
is sensitive to both the number of stars and the IMF, but we adjust the
number of stars so that all models match the observed rms
of the broad-band fluctuations.\footnote{The default, Salpeter, model
has $2.0\times 10^7$ stars per resolution element, the $x=3$ model
has $4.2\times 10^7$ stars, and the bottom-light
model has $0.14\times
10^7$ stars.} In the regime of small
fluctuations the trends in Fig.\ \ref{moddata3.fig}  depend on
the luminosity -- temperature relation on the upper giant branch, and
are independent of the number of low mass stars.
Although there is no direct IMF dependence,
the results presented in this paper are relevant to measurements
of the IMF from integated light, as these rely
on accurate modeling of the contribution of luminous giants to
the spectra of early-type galaxies.

%\subsection{Comparison to Integrated-light Spectra}
%
%TO DO: COMPARE CONSTRAINTS FROM SBF TO VARIATION IN INTEGRATED LIGHT. OR FORGET
%ABOUT IT? MAYBE AFTER REFEREE COMMENTS?

\subsection{Predictions For Large (Negative) Fluctuations}
\label{large.sec}

In this paper we focused on a regime
of relatively small surface brightness fluctuations, of $\pm 5$\,\%.
This range is dictated by the currently-available data: the Poisson
errors quickly increase for fluctuations $<0.95$ and $>1.05$, as we
simply do not have many pixels with variations that large.
As a result of the small observed fluctuation range there are,
on average, $\sim 90$ stars with $L_{\rm bol}>1000$\,L$_{\odot}$
even in the lowest fluctuation pixels. The fluctuation spectra
are therefore completely dominated by the TiO and H$_2$O
features that are characteristic of
these luminous red stars.

Here we briefly consider what we can expect in a regime where there
are far fewer stars per pixel, so that the surface brightness flucations
are much larger. There is a wealth of information in the fluctuation
spectra in this regime, as there are sightlines that are
free of luminous giants. Those pixels, with surface brightness
fluctuations much smaller than 1, provide a window
of the lower giant branch
and the main sequence that is unattainable in integrated light observations.

We simulated fluctuation spectra in this regime, using our default model
with $\langle N_{\rm stars}\rangle = 10^5$ per pixel rather than
$2\times 10^7$. In this model there are pixels that
contain no stars with $L_{\rm bol}
>1000$\,L$_{\odot}$.
The distribution of fluctuations in this model is shown in the
top left panel of Fig.\ \ref{largefluc.fig}. The $1\sigma$ spread
is $\approx 0.22$ but the distribution is asymmetric, with
the high fluctuation tail containing multiple luminous giants.
In the top right panel of Fig.\ \ref{largefluc.fig} we show
fluctuation spectra in five regimes, going from 
surface brightness
fluctuations much lower than 1 to fluctuations much greater than 1.
The fluctuation spectra are
dramatically different in these regimes, as they probe different
types of stars. Low fluctuations (0.7/0.6) probe the variation of
subgiants and low luminosity giants
with temperature, and as these stars have no strong TiO features
the fluctuation spectra are dominated by Na\,I and Ca\,II.
Intermediate fluctuations (1.1/1.0) are predicted to show very strong
TiO features
in the fluctuation spectra, as the amplitude of their broad-band
fluctuation depends on whether they have
zero or one star with $L_{\rm bol}>1000$\,L$_{\odot}$.
The regime of strongly positive fluctuations
(1.5/1.4) resembles the regime probed in our observations
of \targ, that is, the variation of the spectra of very luminous
giants with temperature.

The bottom panels of Fig.\ \ref{largefluc.fig} show the predicted
relations between ramp filter indices and the F814W surface brightness
fluctuation. The relations are not linear, and they depend on the
IMF. Bottom-heavy IMFs with $x=3$ show characteristic asymmetric
trends, with little dependence on the
surface brightness amplitude for SBF\,$<$\,1 and steep relations
for SBF\,$>$\,1. Note that the horizontal and vertical ranges of these
panels are much larger than in the previous sections.

\section{Summary and Conclusions}
\label{conclusions.sec}

In this paper we introduced a new technique for studying old,
``semi-resolved'' stellar populations. We have shown that it is possible to
identify individual pixels, or sightlines, in moderately deep HST imaging
that have an increased
or decreased number of giant stars due to Poisson variations.
By averaging carefully matched
narrow-band filter observations in bins of surface brightness fluctuation,
we have quantified how the spectrum of \targ\ changes when
giant stars are added or taken away. 

The resulting relations between narrow-band indices and broad-band
surface brightness fluctuations are remarkably well reproduced with the
basic ingredients of the
{Conroy} \& {van Dokkum} (2012a) stellar population synthesis model.
The residuals from the predicted relations are very small at
0.03\,\% -- 0.09\,\%. Qualitatively, this means that the model
passes an important test in a previously unexplored regime.
It also means that we know what
the most luminous stars in elliptical galaxies are: giants with
temperatures $T_{\rm eff} \sim 3400$\,K and luminosities $L_{\rm bol}
\sim 1300$\,L$_{\odot}$. Their spectra are well understood, even to the extent
that our default model with the \targ\ abundance pattern is a significantly
better match to the data than a ``standard'' $\alpha-$enhanced model
with [$\alpha/$Fe]\,=\,0.3.
Understanding stars with $L_{\rm bol}>1000$\,L$_{\odot}$
is important as they comprise $\approx 17$\,\%
of the bolometric luminosity of old stellar populations, and
their strong TiO spectral features coincide with the IMF-sensitive features
Na\,I and FeH.  We stress, however, that our data do not
provide information on lower luminosity stars, or on the absolute
number of luminous giants (see \S\,5.1.3).

%Perhaps trivially, given the extensive literature on the topic
%of surface brightness fluctuations, we have also demonstrated
%that these fluctuations
%are, in fact, caused by variations in the number of luminous giants
%(Fig.\ \ref{moddata1.fig}).

We have tested various
perturbations of our default model. The most important result is
that
we can quantitatively constrain the contribution of late M giants
to the integrated light. Such stars are not expected to exist in
substantal numbers in
old stellar populations, but their presence cannot easily
be ruled out
based on integrated light analysis. Importantly,
leaving the contribution of these stars as a free parameter
in fitting stellar populations can lead to degeneracies with the
derived stellar initial mass function, as several TiO bands
overlap with IMF-sensitive features (see \S\,5.2 in {Conroy} \& {van Dokkum} 2012b).
Here we show that, for this particular galaxy,
they cannot contribute more than 3\,\%
to the integrated light.

Taking the stellar templates and isochrones as a given, we have explored
the dependence of the observed ramp filter data
on stellar population parameters. In the regime of small
fluctuations the data are insensitive to the
IMF, but they provide constraints on metallicity and age.
In particular, the fluctuation spectra are
quite sensitive to age, as the number of
late M giants with strong TiO features
drops sharply at very old ages. It is generally difficult
to distinguish stellar populations that are between 10 and
13 Gyr old, and fluctuation spectroscopy is probably among the most
sensitive diagnostics in this regime. 

%There are several limitations to this technique. Galaxies have to
%have intrinsically-smooth stellar populations and no star formation
%or dust. Furthermore, it is technically challenging,
%as the spatial resolution of the {\em Hubble Space Telescope}
%(HST) is required for sufficient dynamic range in the fluctuations.

The analysis in this paper can be extended in various directions.
The two most obvious ones are to obtain actual spectra rather than
narrow-band filter data, and to probe regimes of larger surface
brightness fluctuations. As illustrated in Fig.\ \ref{largefluc.fig}
we can expect qualitative changes in the spectra of fluctuations
$\ll 1$, as these sightlines
offer a relatively unobstructed view of subgiants
and the main sequence.\footnote{This
is conceptually similar to the suggestion by
{Mould} (2012) to mask individual stars in 30\,m telescope
observations of the outer parts of galaxies.}
The analysis could center on differential techniques such as
those employed in this paper, or on direct
modeling of the spectra of
very negative fluctuations.
In principle, it should be possible
to empirically, and spectroscopically,
reconstruct the entire upper RGB of elliptical
galaxies -- something that is impossible to do in any other way.
Among other applications, this
type of analysis should provide strong constraints
on the IMF (bottom panel of Fig.\ \ref{largefluc.fig}).
Observations with integral-field spectrographs on existing
telescopes, particularly when combined with adaptive optics, could
provide this kind of information for galaxies within $\sim 3$\,Mpc.
With JWST and 30m class
telescopes on the ground these studies could be extended to the
Virgo cluster and beyond.

Progress can also be made by combining constraints from fluctuation spectroscopy
with those from fits to integrated-light spectra. The methods
provide complementary information, as traditional blue
spectra are mostly sensitive to stars near the main sequence turn-off.
We note here
that none of the model variations
discussed in this paper lead to dramatic (that is, larger than a few
percent) variations in the integrated-light spectra, with the exception of
the models with very young ages.  Finally, the stellar library can be
augmented. We currently use only 70 distinct templates in our
default library, and in some regimes incomplete coverage leads to
large ``jumps'' from one template to the other. Future
improvements will be enabled by larger and more comprehensive
stellar  libraries.

\begin{acknowledgements}
We thank the anonymous referee for a constructive report that improved
the clarity of the paper.
Support from STScI grant GO-12523 is gratefully acknowledged.
\end{acknowledgements}

%% --------------------------------------------------------------------
%% Fri Aug  1 10:15:07 2014
%%   This file was generated automagically from the files
%%   sbf.bbl and sbf.tex using
%%     nat2jour.pl
%%   This file should accompany sbf-aas.tex.
%% --------------------------------------------------------------------


\begin{references}
\reference{} {Beers}, T.~C., {Flynn}, K., \& {Gebhardt}, K. 1990, \aj, 100, 32
\reference{} {Blakeslee}, J.~P., {Cantiello}, M., {Mei}, S., {C{\^o}t{\'e}}, P., {Barber  DeGraaff}, R., {Ferrarese}, L., {Jord{\'a}n}, A., {Peng}, E.~W., {et al.} 2010, \apj, 724, 657
\reference{} {Blakeslee}, J.~P., {Vazdekis}, A., \& {Ajhar}, E.~A. 2001, \mnras, 320, 193
\reference{} {Bruzual}, G. \& {Charlot}, S. 2003, \mnras, 344, 1000
\reference{} {Caldwell}, N., {Schiavon}, R., {Morrison}, H., {Rose}, J.~A., \& {Harding}, P.  2011, \aj, 141, 61
\reference{} {Conroy}, C. 2013, \araa, 51, 393
\reference{} {Conroy}, C., {Graves}, G.~J., \& {van Dokkum}, P.~G. 2014, \apj, 780, 33
\reference{} {Conroy}, C., {Gunn}, J.~E., \& {White}, M. 2009, \apj, 699, 486
\reference{} {Conroy}, C. \& {van Dokkum}, P. 2012a, \apj, 747, 69
\reference{} {Conroy}, C. \& {van Dokkum}, P.~G. 2012b, \apj, 760, 71
\reference{} {Croton}, D.~J., {Springel}, V., {White}, S.~D.~M., {De Lucia}, G., {Frenk},  C.~S., {Gao}, L., {Jenkins}, A., {Kauffmann}, G., {et al.} 2006, \mnras, 365, 11
\reference{} {Dalcanton}, J.~J., {Williams}, B.~F., {Seth}, A.~C.,
{Dolphin}, A., {Holtzman}, J., {Rosema}, K., {Skillman}, E.~D., 
{Cole}, A., {et al.} 2009, \apjs, 183, 67
\reference{} {Girardi}, L., {Williams}, B.~F., {Gilbert}, K.~M.,
{Rosenfield}, P., {Dalcanton}, J.~J., {Marigo}, P., {Boyer}, M.~L.,
{Dolphin}, A., {et al.} 2010, \apj, 724, 1030
\reference{} {Jensen}, J.~B., {Tonry}, J.~L., {Barris}, B.~J., {Thompson}, R.~I., {Liu},  M.~C., {Rieke}, M.~J., {Ajhar}, E.~A., \& {Blakeslee}, J.~P. 2003, \apj, 583,  712
\reference{} {Krumholz}, M.~R. 2011, \apj, 743, 110
\reference{} {Lan{\c c}on}, A. \& {Wood}, P.~R. 2000, \aaps, 146, 217
\reference{} {Liu}, M.~C., {Graham}, J.~R., \& {Charlot}, S. 2002, \apj, 564, 216
\reference{} {Maraston}, C. 2005, \mnras, 362, 799
\reference{} {Mei}, S., {Blakeslee}, J.~P., {Tonry}, J.~L., {Jord{\'a}n}, A., {Peng}, E.~W.,  {C{\^o}t{\'e}}, P., {Ferrarese}, L., {West}, M.~J., {et al.} 2005, \apj, 625, 121
\reference{} {Mei}, S., {Blakeslee}, J.~P., {C{\^o}t{\'e}}, P., {West}, M.~J.,
{Ferrarese}, L.,  {Jord{\'a}n}, A., {Peng}, E.~W., {Anthony}, A., \& {Merritt}, D. 2007, \apj, 655, 144
\reference{} {Mitchell}, P.~D., {Lacey}, C.~G., {Baugh}, C.~M., \& {Cole}, S. 2013, \mnras,  435, 87
\reference{} {Mould}, J. 2012, \apjl, 755, L14
\reference{} {Nelson}, E., {van Dokkum}, P., {Franx}, M., {Brammer}, G.,
{Momcheva}, I., {F\o{}rster Schreiber}, N., {da Cunha}, E., {Tacconi}, L.,
{et al.} 2014, Nature, 513, 394
\reference{} {No{\"e}l}, N.~E.~D., {Greggio}, L., {Renzini}, A., {Carollo}, C.~M., \&  {Maraston}, C. 2013, \apj, 772, 58
\reference{} {Rayner}, J.~T., {Cushing}, M.~C., \& {Vacca}, W.~D. 2009, \apjs, 185, 289
\reference{} {Salpeter}, E.~E. 1955, \apj, 121, 161
\reference{} {Schiavon}, R.~P. 2007, \apjs, 171, 146
\reference{} {Thomas}, D., {Johansson}, J., \& {Maraston}, C. 2011, \mnras, 412, 2199
\reference{} {Tonry}, J. \& {Schneider}, D.~P. 1988, \aj, 96, 807
\reference{} {Tonry}, J.~L., {Dressler}, A., {Blakeslee}, J.~P., {Ajhar}, E.~A., {Fletcher},  A.~B., {Luppino}, G.~A., {Metzger}, M.~R., \& {Moore}, C.~B. 2001, \apj, 546,  681
\reference{} {Valenti}, J.~A., {Piskunov}, N., \& {Johns-Krull}, C.~M. 1998, \apj, 498, 851
\reference{} {van der Marel}, R.~P., {Cretton}, N., {de Zeeuw}, P.~T., \& {Rix}, H.-W. 1998,  \apj, 493, 613
\reference{} {van Dokkum}, P.~G. 2008, \apj, 674, 29
\reference{} {van Dokkum}, P.~G. \& {Conroy}, C. 2010, \nat, 468, 940
\reference{} ---. 2011, \apjl, 735, L13
\reference{} {van Dokkum}, P.~G., {Bezanson}, R., {van der Wel}, A.,
{Nelson}, E.~J., {Momcheva}, I., {Skelton}, R.~E., {Whitaker}, K.~E.,
{Brammer}, G., {et al.} 2014, \apj, 791, 45
\reference{} {Vazdekis}, A., {Peletier}, R.~F., {Beckman}, J.~E., \& {Casuso}, E. 1997,  \apjs, 111, 203
\reference{} {Walcher}, J., {Groves}, B., {Budav{\'a}ri}, T., \& {Dale}, D. 2011, \apss,  331, 1
\reference{} {Wing}, R.~F. \& {Ford}, Jr., W.~K. 1969, \pasp, 81, 527
\reference{} {Worthey}, G. 1994, \apjs, 95, 107
\reference{} {Worthey}, G., {Faber}, S.~M., \& {Gonzalez}, J.~J. 1992, \apj, 398, 69
\end{references}
\end{document}